\documentclass[jcp,aps,twocolumn,showpacs,superscriptaddress,epsfig,amsmath,amssymb,eqsecnum]{revtex4-1}
\usepackage{epsfig,amsmath,amssymb,bm,epsf,graphicx,psfrag,verbatim,subfigure,framed,dcolumn,ifpdf,hyperref}
\usepackage[usenames,dvipsnames]{xcolor}
\usepackage{bbm}
\usepackage{mathrsfs}
\usepackage{amsfonts}
\usepackage{color}
\usepackage{pbox}
\usepackage{soul} 
\def\mon{a}

\def\rod-like{\mathfrak{K}}

\def\llangle{\langle\!\langle}
\def\rrangle{\rangle\!\rangle}
\newcommand{\be}{\begin{equation}}
\newcommand{\ee}{\end{equation}}
\newcommand{\ba}{\begin{eqnarray}}
\newcommand{\ea}{\end{eqnarray}}
\newcommand{\bw}{\begin{widetext}}
\newcommand{\ew}{\end{widetext}}

\newcommand{\Pv}{{\bm{P}}}

\newcommand{\rv}{{\bm{r}}}

\newcommand{\cv}{{\bm{c}}}

\newcommand{\pv}{{\bm{p}}}

\newcommand{\uv}{{\bm{u}}}

\newcommand{\Rv}{{\bm{R}}}
\newcommand{\rh}{{\widehat{\bm{r}}}}

\newcommand{\tv}{\bm{t}}

\definecolor{darkblue}{rgb}{0.0, 0.0, 0.55}

\begin{document}

\title{Molecular recognition by van der Waals interaction between polymers with sequence-specific polarizabilities}
\author{Bing-Sui Lu}
\thanks{Email: \texttt{bing-sui.lu@fmf.uni-lj.si}}
\affiliation{Department of Physics, Faculty of Mathematics and Physics, University of Ljubljana, Jadranska ulica 19, SI-1000 Ljubljana, Slovenia}
\author{Ali Naji}
\affiliation{School of Physics, Institute for Research in Fundamental Sciences (IPM), P.O. Box 19395-5531, Tehran, Iran}
\author{Rudolf Podgornik}
\affiliation{Department of Physics, Faculty of Mathematics and Physics, University of Ljubljana, Jadranska ulica 19, SI-1000 Ljubljana, Slovenia}

\begin{abstract}
We analyze van der Waals interactions between two rigid polymers with sequence-specific, anisotropic polarizabilities along the polymer backbones, so that the dipole moments fluctuate parallel to the polymer backbones. Assuming that each polymer has a quenched-in polarizability sequence which reflects, for example, the polynucleotide sequence of a double-stranded DNA molecule, we study the van der Waals  interaction energy between a pair of such polymers with rod-like structure for the cases where their respective polarizability sequences are (i)~distinct and (ii)~identical, with both zero and non-zero correlation length of the polarizability correlator along the polymer backbones in the latter case. For identical polymers, we find a novel $r^{-5}$ scaling behavior of the van der Waals interaction energy for small inter-polymer separation $r$, in contradistinction to the $r^{-4}$ scaling behavior of distinct polymers, with furthermore a pronounced angular dependence favoring attraction between sufficiently aligned identical polymers.  Such behavior can assist the molecular recognition between polymers.  
\end{abstract}


\maketitle

\section{Introduction}
 
Van der Waals (vdW) forces are ubiquitous in Nature, being especially important in the nanoscale world \cite{French}. They arise owing to the presence of fluctuating permanent and/or induced dipoles on atoms, molecules and molecular aggregates. Dipolar fluctuations enter the calculation of the vdW interaction (free) energies through the corresponding frequency-dependent and, in general, anisotropic dielectric response functions \cite{Dalvit, Bordag, Parsegian}. The latter case is particularly germane for long chain-like molecules, where the dielectric response functions along the molecular chain axis and perpendicular to it can differ significantly. A good example would be the strong optical anisotropy of DNA \cite{Macnot} or long carbon nanotubes \cite{Rajter} with significantly different dielectric response functions in the axial and the radial directions. The role of dielectric anisotropy in the context of polymers was recently recognized specifically in the context of liquid crystalline ordering in the case of anisotropic polarizable polymers~\cite{dean_podgornik}, leading to important consequences. In general, vdW interactions between anisotropic media lead not only to long-range interactions but also to long-range torques as has been recognized some time ago \cite{Capasso,Rajter}. The details of this anisotropy effect are in fact many faceted and are discernible in the non-retarded as well as in the retarded limit of vdW interactions \cite{Siber}. 

Anisotropy is not the only important peculiarity of polymer polarizability. It is now recognized, based on extensive {\em ab initio} calculations and UV-vis molar absorbance measurements, that the specificity of the oligonucleotide sequence of DNA molecules strongly influences the dielectric response of the molecule and its polarizability properties~\cite{schimelman}.  The question thus arises whether the detailed sequence of monomer polarizabilities in heteropolymers would influence in some essential capacity the interaction between two identical polymers, enabling in this way the interacting biological systems to recognize at a finite separation the polarizability pattern, leading to {\sl molecular recognition}. Similar types of molecular recognition problems have been studied in the context of electrostatic interactions between randomly, irreversibly charged objects \cite{panyukov_rabin,disorder-PRL,jcp2010,pre2011,epje2012,jcp2012,speake,Ben-Yaakov-dis} and randomly patterned surfaces interacting through a generic finite range interaction potential, as a prototype model of protein-protein interactions \cite{lukatsky1,lukatsky2}. It seems, however, that at least for polynucleotides, the sequence specificity is much more pronounced in terms of polarizability than charge distribution and thus there is some interest to investigate the sequence-specific polarizability effects on vdW interactions.

For our purpose we consider a pair of \emph{rigid}, \emph{charge neutral} and \emph{polarizable} polymers (or, more precisely, heteropolymers), which experience induced dipole--induced dipole interactions between monomers characterized by different values of polarizability. 
In general, the polarizability is a tensor, with different radial and axial components (the axial direction being parallel to the backbone of the polymer). For long and thin polymers, the polarization of the polymer is mostly concentrated along its backbone, and one can then neglect the radial contribution to the polarizability, which is the approximation that we will consider in this paper. The local polarizability of a heteropolymer depends on the identity of the particular monomer (e.g., base pairs along the double-stranded (ds) DNA molecule), and we can thus take it to be set irreversibly, i.e., \emph{quenched}. For simplicity, we consider only the case of non-retarded interactions since the molecular recognition effects are expected to operate mainly on short length scales. Furthermore, we consider the regime of high temperatures and focus on the contribution of static polarizabilities, which are nevertheless anisotropic so that the molecular dipoles can not be considered as freely rotating \cite{ksm}. Quite generally, in aqueous environment the static term contributes more then 50 \% to the total vdW interaction potential \cite{Parsegian}. Finally, we delimit our analysis to the case of \emph{rod-like} polymers, which we will study using the so-called shish-kebab model~\cite{doi-edwards}; this model should also be appropriate for dsDNA molecules at small separations. 

We shall derive a field theory that accounts for thermal fluctuations of induced dipoles along the polymer backbone, which we shall then apply to study the effective interaction energy between a pair of rod-like polymers with  \emph{distinct} (i.e., completely uncorrelated) polarizability sequences or a pair of polymers with \emph{identical} (i.e., maximally correlated) polarizability sequences.  By comparing the vdW interaction energies of a pair of distinct polymers and a pair of  identical polymers, we find the surprising result that the interaction energies display fundamentally different behaviors when the inter-polymer separation $r$ is smaller than the length of each polymer. The sequence-averaged (in the sense to be defined in Sec.~\ref{sec:sequence}) interaction energy for a pair of distinct polymers shows a $r^{-4}$ dependence and is invariant with respect to the inversion of either polymer about its center of mass, whereas the sequence-averaged interaction energy for two identical polymers has a $r^{-5}$ scaling form and is non-degenerate with respect to polymer inversion. The latter effect leads to favoring the selection of pairs of identical over distinct rod-like polymers and, moreover, amongst those pairs of identical polymers selected, there is a higher probability that these polymers are (i)~parallel or (ii)~anti-parallel with the line joining their centers of mass parallel to the tangent vector of each. Such features can assist in the recognition of molecular sequences, such as in DNA molecules, interacting at short distances. 
 
Here is a brief outline of the paper. In Sec.~II, we describe our model of rigid, polarizable and charge neutral  polymers interacting via induced dipoles. We make an interesting observation about the character of interactions between induced dipoles along charge neutral \emph{semi-flexible} polymers, and present a discussion of sequence-specific polarizabilities, describing two models for the correlation between such polarizabilities. In Sec.~III, we derive a field theory for the interactions of polymers via induced dipoles fluctuations. In Sec.~IV, we present our results for the effective interaction energy between a pair of rigid, rod-like polymers. 
In Sec.~V, we give a summary and discussion of the key results. 

\section{Model description of rigid polarizable polymers}

Let us denote the position vector of the $m$-th monomer (which can be, for instance, a nucleotide base-pair of a dsDNA molecule) on the $i$-th polymer by the symbol $\Rv^{(i)}(m)$. We assume that the polymer has $M$ monomers (so $m=1,\ldots,M$) each of equal length $\mon$, and there are a total of $N$ polymers (so $i=1, \ldots, N$). Each segment comes equipped with a thermally fluctuating, instantaneous dipole moment represented by a vector $\pv^{(i)}(m)$ and a polarizability $\alpha^{(i)}(m)$. We assume that the $\alpha^{(i)}(m)$ distribution along the polymer backbone is quenched, i.e., set irreversibly for a given realization of the polymer. We assume that there are no permanent dipole moments and all the dipoles are induced; thus $\pv^{(i)}(m)$ is a quantity fluctuating about zero. 

For such a system, the Hamiltonian is given by 
\ba
&&H \! = \! \frac{1}{2}\sum_{ij} {\sum_{mn}}^\prime 
\pv^{(i)}(m)  \!\cdot \!
 \frac{\partial^2 G(\Rv^{(i)}(m), \Rv^{(j)}(n))}{\partial \Rv^{(i)}(m) \partial \Rv^{(j)}(n)} \!\cdot\!  \pv^{(j)}(n)
\nonumber\\
&&\quad\quad+ \frac{1}{2} \sum_i \sum_m \frac{(\pv^{(i)}(m))^2}{\alpha^{(i)}(m)},  
\label{eq:Hamiltonian}
\ea 
where the prime indicates that the terms with $i=j$ and $m=n$ are omitted from the first term. 
The above expression is an analogue of the polarizable field model proposed in Ref.~\cite{dean_polarizable,dean_polarizable1,dean_polarizable2}, developed here for the case of heteropolymers. 
The first term in Eq. (\ref{eq:Hamiltonian}) describes the electrostatic interaction between a pair of induced dipole moments, and the second term describes the self-polarization of the polymer, which can be regarded as the elastic energy of the local charge distortion~\cite{frydel1,frydel2}. Here, $G(\rv_1, \rv_2)$ denotes the Green's function, defined by 
\be
G(\rv_1,\rv_2) = \frac{1}{4\pi\epsilon\epsilon_0|\rv_1-\rv_2|}, 
\label{eq:green_function}
\ee
where $\epsilon$ is the relative permittivity of the dielectric medium the polymers inhabit. If we assume that water can be approximated as a uniform dielectric medium then $\epsilon \simeq 80$ for the case of  aqueous solutions  at room temperature $T=293$~K. 

\subsection{Interaction between effective monopoles}

Here we make an interesting observation that the first term of Eq.~(\ref{eq:Hamiltonian}) can be expressed in a form that describes interactions between effective monopolar fluctuations, if (i)~the polymers are semiflexible and (ii)~the dipole fluctuations are parallel to the polymer backbone. Taking the continuum limit, i.e., $m \rightarrow s/\mon$, where $-\ell/2 \leq s \leq \ell/2$ is the continuous arc-length coordinate, these two conditions are summarized by the equation:
\be
\pv(s) = p(s)\tv(s) = p(s) \dot{\Rv}(s). 
\label{eq:cont}
\ee
By defining the total polarization vector $\Pv(\rv)$, where
\be
\Pv(\rv) 
= \frac{1}{\mon}\sum_{i=1}^{N}\int \! ds\,p^{(i)}(s)\dot{\Rv}^{(i)}(s)\delta(\rv-\Rv^{(i)}(s)), 
\ee
We can rewrite the first term of Eq.~(\ref{eq:Hamiltonian}) as follows: 
\ba
&&\frac{1}{2}\sum_{ij} {\sum_{mn}}^\prime \pv^{(i)}(m)  \!\cdot \!
 \frac{\partial^2 G(\Rv^{(i)}(m), \Rv^{(j)}(n))}{\partial \Rv^{(i)}(m) \partial \Rv^{(j)}(n)} \!\cdot\!  \pv^{(j)}(n)
 \nonumber\\
 && \rightarrow
\frac{1}{2}\int\! \!\!\int d\rv d{\rv'} \, 
\Pv(\rv)\!\cdot\! \nabla_\rv \nabla_{\rv'}G(\rv -\rv')\!\cdot\! \Pv(\rv'). 
\label{eq:V_dipole1}
\ea
By partial integration, the expression~(\ref{eq:V_dipole1}) simplifies to  
\ba
&&\frac{1}{2} \! \int\! \!\!\int d\rv d{\rv'} \,
\nabla_\rv \!\cdot\! \Pv(\rv) G(\rv -\rv') \nabla_{\rv'} \!\cdot\! \Pv(\rv'), 
\ea
where $\nabla_{\rv}\cdot\Pv(\rv)$ is the polarization monopolar charge density:  
\ba
&&\nabla_{\rv} \!\cdot\! \Pv(\rv) 
= \frac{1}{\mon}\sum_{{i}=1}^{N}\!\int \!ds\, p^{(i)}(s) \frac{\partial \Rv^{(i)}(s)}{\partial s} \!\cdot\! 
\nabla_\rv\delta(\rv - \Rv^{(i)}(s))
\nonumber\\ 
&&\quad= -\frac{1}{\mon}\sum_{{i}=1}^{N}\int ds\, p^{(i)}(s) \partial_s \delta(\rv - \Rv^{(i)}(s))
 \nonumber\\
 &&\quad= \frac{1}{\mon}\sum_{{i}=1}^{N} \bigg[ p^{(i)}(-\frac{\ell}{2})\,\delta(\rv-\Rv^{(i)}(-\frac{\ell}{2})) 
 \nonumber\\
 &&\quad\quad\quad\quad\quad
 - p^{(i)}(\frac{\ell}{2})\,\delta(\rv-\Rv^{(i)}(\frac{\ell}{2})) 
 \nonumber\\
 &&\quad\quad\quad\quad\quad
 + \int ds\,\dot{p}^{(i)}(s)\,\delta(\rv - \Rv^{(i)}(s)) \bigg].
 \label{eq:div_P}
\ea
There is a positive polarization charge of magnitude $p_{i}(-\frac{\ell}{2})/a$ at $\Rv_{i}(-\frac{\ell}{2})$, a negative polarization charge of $p_{i}(-\frac{\ell}{2})/a$ at $\Rv_{i}(\frac{\ell}{2})$, and a linear polarization charge density of $\dot{p}_{i}(s)/a$ along the chain. 
Thus, for dipole fluctuations that are induced parallel to the polymer backbones, their interaction kernels can be transformed to monopolar Coulomb potentials, and their interaction can be regarded as that between effective monopolar fluctuations. 

\subsection{Quenched sequence-specific polarizabilities}
\label{sec:sequence} 

\begin{figure}
		\includegraphics[width=0.40\textwidth]{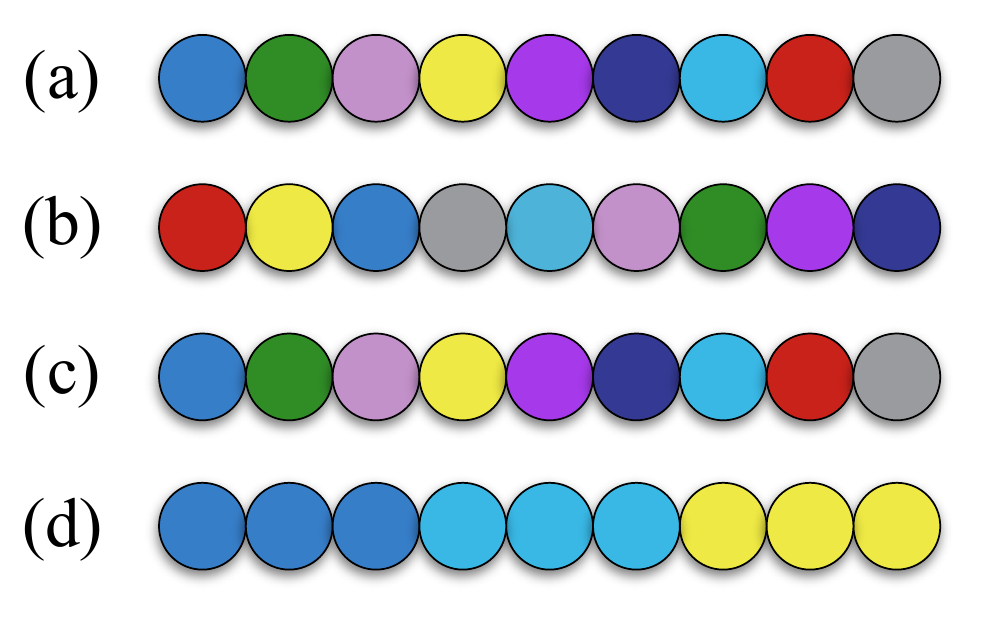}
	\caption{Examples of sequences and sequence pairs that are studied in Sec.~IV. Each colored circle represents a monomer segment, and distinct colors represent distinct polarizabilities. Sequences~(a) and (b) are distinct, and their interaction free energy is studied in Sec.~IV A, whereas sequences~(a) and (c) are identical, and studied in Sec.~IV B. Polarizabilities on sequences~(a), (b) and (c) have zero intra-chain sequence correlation length, whilst those on sequence~(d) have nonzero intra-chain sequence correlation length. The effect of nonzero correlation length is studied in Sec.~IV C.} 
\label{fig:sequences}
\end{figure}

We now address the issue of  sequence specificity for rigid polymers. In the continuum limit with $m \rightarrow s/\mon$ (see above), a polymer $i$ with a chosen sequence of monomers has also a quenched-in polarizability sequence $\alpha^{(i)}(s)$, where $\alpha^{(i)}(s)$ of each monomer along the polymer chain $i$ depends on its intrinsic physical properties. We need not specify at this point what the monomer polarizability is due to or indeed how the monomers are defined. Assuming that the different monomers have a mean polarizability, which we call $\alpha^{(i)}_0$, then the differences between monomers are reflected in the deviation $\delta\alpha^{(i)}(s)= \alpha^{(i)}(s) - \alpha^{(i)}_0$ of the polarizability of each monomer segment from the mean polarizability value. 

In the absence of knowledge about how polarizabilities $\alpha^{(i)}(s)$ are distributed along each polymer $i$, we shall assume, for simplicity, that the values of the polarizabilities are drawn from a Gaussian probability distribution $P_\alpha[\alpha^{(i)}(s)]$, defined by its average and its variance. Denoting the {\em sequence average} over $P_\alpha$ as $\llangle \ldots \rrangle$, the average of polarizability is then
\be
\llangle \alpha^{(i)}(s) \rrangle = \alpha_0,
\ee
and the {\em sequence correlator} is defined as
\be
\llangle \delta\alpha^{(i)}(s)\, \delta\alpha^{(j)}(s') \rrangle \equiv g^2 \mon\,\delta(s-s')\delta_{ij}.
\label{equ:gfyuerw}
\ee
On the other hand, if the monomeric polarizabilities along the polymer backbone are  correlated over a nonzero length scale, we can approximate the sequence correlator by  
\be
\llangle \delta\alpha^{(i)}(s)\, \delta\alpha^{(j)}(s') \rrangle \simeq {g^2 \mon\,} f(s - s')\delta_{ij},
\label{equ:beruwy}
\ee
where $f(s - s')$ can be any decaying function of the argument such as a Gaussian, viz., $f(s - s') \equiv ({\sqrt{2\pi}\sigma})^{-1} e^{-\frac{(s-s')^2}{2\sigma^2}}$, where $\sigma$ is a typical length scale. Other choices are of course also possible. The form Eq.~(\ref{equ:gfyuerw}) thus corresponds to a vanishing correlation length.

In what follows we will be interested in the following three cases pertinent to two ($i, j = 1,2$) interacting polymer chains [cf. Fig.~\ref{fig:sequences}]: 
\begin{enumerate}
\item[i)]  Polymers are \emph{distinct}: $\alpha^{(1)}(s) \neq \alpha^{(2)}(s)$. 
There is a \emph{different} probability distribution $P_\alpha[\alpha^{(i)}(s)]$ corresponding to \emph{each} polymer, and polarizabilities on distinct polymers are uncorrelated. Thus, 
we have
\be
\llangle \alpha^{(1)}(s) \,\alpha^{(2)}(s') \rrangle = \alpha_0^2.  
\ee
\item[ii)] The pair of polymers are \emph{identical}, i.e., $\alpha^{(1)}(s) = \alpha^{(2)}(s) \equiv \alpha(s)$. 
In this case, there is the \emph{same} probability distribution $P_\alpha[\alpha(s)]$ of the polarizabilities for \emph{either} polymer. If we assume a vanishing correlation length, we obtain the following polarizability correlator on each polymer:  
\be
\llangle \alpha(s)\,\alpha(s') \rrangle = \alpha_0^2 + {g^2 \mon} \, \delta(s-s').  
\label{equ:aa_ii}
\ee
\item[iii)]  The third case is similar to the second case (identical polymers), except that the intra-chain correlation length of the sequence polarizability pattern is now nonzero. In this case, we have 
\be
\llangle \alpha(s)\,\alpha(s') \rrangle = \alpha_0^2 + {g^2 \mon} f(s - s'),  
\ee
where again $ f(s - s')$ can be any decaying function of the argument. 
\end{enumerate}
Next, we will derive a field theoretical  representation for the partition function of the system, which will enable us to study the behavior of the sequence-averaged vdW interaction free energy for the above three cases for a pair of rigid, rod-like polymers. 

\section{Field theory for rigid polymers interacting via induced dipoles}

We are interested in evaluating the effective interaction (free) energy between a pair of rigid, polarizable polymers, where the dipole fluctuations have been integrated out. The task is made non-trivial by virtue of the fact that the interacting dipoles in the Hamiltonian~(\ref{eq:Hamiltonian}) are coupled to one another. They can be decoupled by introducing an auxiliary field $\varphi$ and performing a Hubbard-Stratonovich transformation on the partition function. The resulting expression for the partition function is then amenable to a systematic approximation procedure using techniques of perturbation theory, which enables one to systematically approximate the effective polymer interaction energy. 

\subsection{Hubbard-Stratonovich transformation}

We can rewrite 
\ba
\label{eq:HS}
&&e^{-\frac{\beta}{2}\sum_{ij} \sum_{mn}^\prime \pv^{(i)}(m)  \cdot 
 \frac{\partial^2 G(\Rv^{(i)}(m), \Rv^{(j)}(n))}{\partial \Rv^{(i)}(m) \partial \Rv^{(j)}(n)} \cdot  \pv^{(j)}(n) }
\\
&=& 
e^{-\frac{1}{2}\ln \det G}\!\!
 \int \! \mathcal{D}\varphi\, e^{-\frac{1}{2}\beta\epsilon\epsilon_0\int \!d\rv\, (\nabla\varphi)^2 
+ i \beta \sum_{i, m} \pv^{(i)} \cdot \nabla\varphi(\Rv^{(i)})},\nonumber
\ea
where $\beta=1/k_{\mathrm{B}}T$, and we recall that the prime denotes the exclusion of terms for which $m=n$ on the same polymer. For a given chain conformation, but allowing for thermal fluctuations of dipole moments, the partition function is given by 
\ba
Z &=& 
\prod_{i,m}\int d\pv^{(i)}(m) \, e^{-\beta H}
\nonumber\\
&=& e^{-\frac{1}{2}\ln \det G} \! \int \! \mathcal{D}\varphi\, e^{-\frac{1}{2}\beta\epsilon\epsilon_0\int\! d\rv \,(\nabla\varphi)^2}
\nonumber\\
&&\times
\prod_{i,m}\int d\pv^{(i)}(m) \, 
e^{-\frac{1}{2}\beta \sum_{i,m} \big( \alpha^{(i)}(m) \big)^{-1} (\pv^{(i)}(m))^{2}} 
\nonumber\\
&&\quad\times
e^{i\beta\sum_{i,m} \pv^{(i)}(m) \cdot \nabla\varphi(\Rv^{(i)}(m))},
\ea
where we have made use of the Hubbard-Stratonovich transformation in Eq. (\ref{eq:HS}). Note that the dipoles are now decoupled, which enables a straightforward integration over the dipole fluctuations. 

For polymers that are much more polarizable along their backbone than perpendicular to it, the induced dipoles prefer to fluctuate parallel to the polymer backbone [cf. Eq. (\ref{eq:cont})] in the direction of the local tangent vector of the polymer, $\tv^{(i)}(m)$.  The dipole orientation is thus fixed by the polymer backbone orientation, and does not fluctuate if the polymer conformation is fixed. On the other hand, the magnitude of the dipole can still fluctuate in the whole accessible interval. Negative values correspond to the dipole pointing in a direction opposite to the tangent vector, and positive values correspond to the dipole pointing in the same direction as the tangent vector. The functional integral over $\pv^{(i)}(m)$ then becomes a functional integral only over the magnitude $p^{(i)}(m)$. This involves evaluating the following product of Gaussian integrals: 
\ba
&&\prod_{i,m}\int dp^{(i)}(m) \, 
e^{-\frac{1}{2}\beta \sum\limits_{i,m} \big( \alpha^{(i)}(m) \big)^{-1} (p^{(i)}(m))^{2}} 
\\
&&\quad\times 
e^{i\beta\sum\limits_{i,m} p^{(i)}(m) \tv^{(i)}(m) \cdot \nabla\varphi(\Rv^{(i)}(m))} 
\nonumber\\
&& = \prod_{i,m}\sqrt{2\pi k_{\mathrm{B}}T \alpha^{(i)}(m)}
e^{-\frac{\beta}{2}\sum\limits_{i,m} \alpha^{(i)}(m) \big( \tv^{(i)}(m) \cdot \nabla\varphi(\Rv^{(i)}(m)) \big)^{2}}
\nonumber
\ea
The local polarizability of the polymer segments thus gives rise to a renormalization of the local dielectric constant. 
The partition function now reads    
\ba
Z &=& Z_0 \! \int \! \mathcal{D}\varphi\, \, e^{-\frac{1}{2} \beta\int\! d\rv \,\epsilon\epsilon_0(\nabla\varphi)^2}
\nonumber\\
& &\times e^{-\frac{1}{2}\beta\sum\limits_{i,m} \alpha^{(i)}(m) \big( \tv^{(i)}(m) \cdot \nabla\varphi(\Rv^{(i)}(m)) \big)^{2}}, 
\label{eq:part_function:2}
\ea
where  we have defined the bare partition function $Z_0$ by 
\be
Z_0 \equiv \! \int \! \mathcal{D}\varphi\, \, e^{-\frac{\beta}{2}\!\int\!d\rv\,d\rv' \, \varphi(\rv)G^{-1}(\rv,\rv')\varphi(\rv')} = e^{-\frac{1}{2}\ln \det G},
\ee
and $G^{-1}(\rv,\rv') \equiv -\epsilon\epsilon_0\nabla_\rv^2\delta(\rv-\rv')$, so that the Boltzmann average with respect to $Z_0$ is given by 
\be
\langle \ldots \rangle_0 \equiv Z_0^{-1} \!\! \int \! \mathcal{D}\varphi\, (\ldots) e^{-\frac{\beta}{2}\!\int\!d\rv\,d\rv' \, \varphi(\rv)G^{-1}(\rv,\rv')\varphi(\rv')}.
\ee
In Eq.~(\ref{eq:part_function:2}) we have subtracted away the constants involving products of $\sqrt{\alpha^{(i)}(m)}$. The meaning of the exponent is as follows. Because there are no fixed charges, the mean electrostatic field is zero. On the other hand, dipole fluctuations give rise to electric field fluctuations $i\nabla\varphi$ in space. Owing to polarizability of the chains, the dielectric permittivity of a segment along a given chain is distinct from that of other segments, and also distinct from the dielectric permittivity of the surrounding space. 

Since the representation~(\ref{eq:part_function:2}) entails a Gaussian integral over auxiliary fields it can be evaluated explicitly, yielding
\ba
\label{eq:part_function:3}
\ln Z &=& \ln Z_0 
\\
&&
-\frac{1}{2}\ln \det \big(
(\epsilon\epsilon_0\nabla^2 + \sum\limits_{i} \sigma^{(i)}_{ab}(\rv) \nabla_a \nabla_b) \delta(\rv-\rv')
\big)
\nonumber
\ea
where we defined $\sigma^{(i)}_{kl}(\rv) = \int ds \, \alpha^{(i)}(s) \, t_a^{(i)}(s) t_b^{(i)}(s) \delta( \rv - \Rv^{(i)}(s))$ in the continuum limit, where the subindices $a, b=1, 2, 3$ denote the Cartesian components. The above expression for the partition function still retains the full dependance on the polarizability sequence $ \alpha^{(i)}(s)$.

\subsection{Perturbation theory}

In order to proceed, we make use of perturbation theory. 
We expand $Z$ to quadratic order in $\alpha^{(i)}$: 
\ba
Z &\simeq& Z_0^{-1} \! \int \! \mathcal{D}\varphi\, \,   
e^{-\frac{1}{2}\beta\int d\rv \, \epsilon\epsilon_0(\nabla\varphi)^2}
\nonumber\\
&&\times
\Big\{  
1 - \frac{1}{2}\beta\sum_{i,m} {\alpha}^{(i)}(m) \big( \tv^{(i)}(m) \cdot \nabla \varphi(\Rv^{(i)}(m)) \big)^2 
\nonumber\\
&&\quad+ \frac{1}{8}\beta^2 \sum_{ij} \sum_{mn}     
{\alpha}^{(i)}(m) \, {\alpha}^{(j)}(n) 
\nonumber\\
&&\quad\quad\times
\big( \tv^{(i)}(m) \cdot \nabla \varphi(\Rv^{(i)}(m)) \big)^2 
\nonumber\\
&&\quad\quad\times
\big( \tv^{(j)}(n) \cdot \nabla \varphi(\Rv^{(j)}(n)) \big)^2 
\Big\}
\label{eq:part_function}
\ea
Following standard techniques of field theory~\cite{zinn-justin}, we find the correlator
\ba
&&\langle \nabla_a^{(i)} \varphi(\Rv^{(i)}(m)) \nabla_b^{(j)} \varphi(\Rv^{(j)}(n)) \rangle_0 
\nonumber\\
&&\qquad\quad=
k_{\mathrm{B}}T\, \frac{\partial^2 G(\Rv^{(i)}(m), \Rv^{(j)}(n))}{\partial R_a^{(i)}(m)\partial R_b^{(j)}(n)}.
\label{eq:correlator1}
\ea
Applying Wick's theorem to $Z$ in Eq.~(\ref{eq:part_function}) and using Eq.~(\ref{eq:correlator1}), we obtain after re-exponentiation \be
Z \simeq e^{-\beta (F_{{\rm self}} + F_{{\rm int}})}.
\label{eq:Z_finite}
\ee
Here, $F_{{\rm self}}$ and $F_{{\rm int}}$ are respectively the free energy of individual polymers and the free energy of interaction between pairs of polymers (for given polymer conformations $\{ \Rv^{(i)}(m), \tv^{(i)}(m) \}$), given by  
\begin{subequations}
\ba
\label{eq:F_self}
&&F_{{\rm self}}[\{ \Rv^{(i)}(m), \tv^{(i)}(m) \}]  
\\
&&= -\frac{k_{\mathrm{B}}T}{4}\sum_i\!\! \sum_{m \neq n} \alpha^{(i)}(m) \alpha^{(i)}(n) 
t_a^{(i)}(m) t_b^{(i)}(m) t_c^{(i)}(n) t_d^{(i)}(n)
\nonumber\\
&&
\quad\quad\times\frac{\partial^2 G(\Rv^{(i)}(m),\Rv^{(i)}(n))}{\partial R_a^{(i)}(m) \partial R_c^{(i)}(n)}
\frac{\partial^2 G(\Rv^{(i)}(m),\Rv^{(i)}(n))}{\partial R_b^{(i)}(m) \partial R_d^{(i)}(n)}
\nonumber\\
&&F_{{\rm int}}[\{ \Rv^{(i)}(m), \tv^{(i)}(m) \}]
\label{eq:F_interaction}
\\
&&= -\frac{k_{\mathrm{B}}T}{4}\sum_{i \neq j}\! \sum_{mn} \alpha^{(i)}(m) \alpha^{(j)}(n) 
t_a^{(i)}(m) t_b^{(i)}(m) t_c^{(j)}(n) t_d^{(j)}(n)
\nonumber\\
&&
\quad\quad\times\frac{\partial^2 G(\Rv^{(i)}(m),\Rv^{(j)}(n))}{\partial R_a^{(i)}(m) \partial R_c^{(j)}(n)}
\frac{\partial^2 G(\Rv^{(i)}(m),\Rv^{(j)}(n))}{\partial R_b^{(i)}(m) \partial R_d^{(j)}(n)}
\nonumber
\ea
\end{subequations}
In deriving Eq.~(\ref{eq:Z_finite}), there are singular terms that involve Coulomb Green's function evaluated with two coincident field points on the same polymer, but their derivatives vanish and thus these terms do not contribute to $Z$. 

Using Eq.~(\ref{eq:green_function}) and again going to the continuum representation, $F_{{\rm int}}$ can be re-written as 
\ba
&&F_{{\rm int}} 
\\
&&
= -\frac{k_{\mathrm{B}}T}{64\pi^2(\epsilon\epsilon_0)^2} \sum_{i \neq j} \int  \! \frac{ds}{\mon}  \! \int  \! \frac{ds'}{\mon}  \,     
{\alpha}^{(i)}(s) \, {\alpha}^{(j)}(s') 
\nonumber\\
&&\quad\times
\left[
\frac{\tv^{(i)}(s)\!\cdot\!\tv^{(j)}(s')}{R_{ij}^3}
-\frac{3(\tv^{(i)}(s)\!\cdot\!\Rv_{ij})(\tv^{(j)}(s')\!\cdot\!\Rv_{ij})}{R_{ij}^5}
\right]^2,
\nonumber
\label{eq:Fint_simple} 
\ea
where $\Rv_{ij} \equiv \Rv^{(i)}(s) - \Rv^{(j)}(s')$. The continuum limit of the interaction free energy is clearly non-pathological for different polymer chains ($i \neq j$). The above form of the interaction energy clearly reduces to the standard  $R_{ij}^{-6}$ form for isotropic chains at large separation. 

\section{Rod-like polarizable polymers}

Let us specialize to the case of two rod-like polymers each of length $\ell$, and consider the three cases delineated above: (i)~distinct polymers, (ii)~identical polymers with zero intra-chain sequence polarizability correlation length, and (iii)~identical polymers with non-zero intra-chain sequence polarizability correlation length.
The $P_\alpha[\alpha^{(i)}(s)]$-averaged (or sequence-averaged) interaction energy for case (i) will be found to be identical to that obtained for polymers each having a uniform polarizability $\alpha_0$, whereas the sequence-averaged interaction energies for cases (ii) and (iii) will depend also on the variance around $\alpha_0$.  

\subsection{Distinct polymers}

After appropriate sequence averaging, the interaction free energy becomes
\ba
\label{eq:fint}
&&\llangle F_{{\rm int}} \rrangle = F_0 
\\
&&
\quad\equiv -\frac{k_{\mathrm{B}}T{\bar{\alpha}}^2}{32\pi^2(\epsilon\epsilon_0)^2} 
\int_{-\ell/2}^{\ell/2} \! ds_1 \! \int_{-\ell/2}^{\ell/2} \! ds_2 
\bigg[
\frac{\tv^{(1)}(s_1)\!\cdot\!\tv^{(2)}(s_2)}{R_{12}^3}
\nonumber\\
&&\quad\quad\quad\quad\quad\quad\quad-\frac{3(\tv^{(1)}(s_1)\!\cdot\!\Rv_{12})(\tv^{(2)}(s_2)\!\cdot\!\Rv_{12})}{R_{12}^5}
\bigg]^2.
\nonumber
\ea
Here, we have defined $F_0$ as the sequence-averaged interaction free energy of a pair of {\em distinct} rod-like polymers, $\bar{\alpha} \equiv \alpha_0/\mon$ is the average polarizability per unit length of each monomer, and we have let $-\ell/2 \leq s_1, s_2 \leq \ell/2$~\cite{footnote_inversion_symmetry}. The pair interaction free energy of distinct polymers is thus equivalent to that of polymers of uniform polarizabilities $\alpha_0$. 
\begin{figure}
		\includegraphics[width=0.35\textwidth]{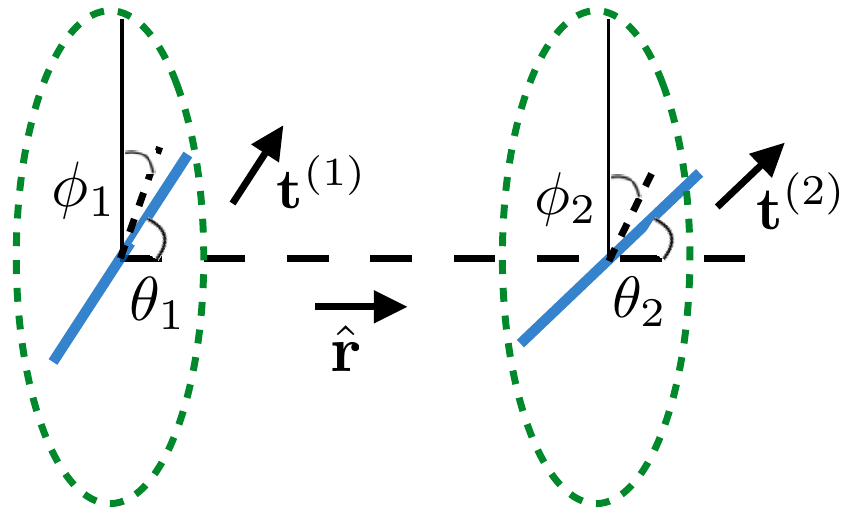}
	\caption{A pair of rod-like polymers parallel to unit vectors $\tv^{(1)}$ and $\tv^{(2)}$ respectively and whose separation runs parallel to the unit vector $\rh$ can equivalently be described in spherical coordinate geometry with $\rh$ as the reference axis, $\theta_1$ and $\theta_2$ as polar angles, and $\phi_1$ and $\phi_2$ as azimuthal angles. The two polymers are co-planar if $\phi_1 = \phi_2$.} 
\label{fig:orientations}
\end{figure}

Equation~(\ref{eq:fint}) can also be expressed in the following form [see App.~\ref{app1}]: 
\ba
&&F_0 = -\frac{k_{\mathrm{B}}T{\bar{\alpha}}^2}{32 \pi^2 (\epsilon\epsilon_0)^2} 
\int_{-\ell/2}^{\ell/2} \! ds_1 \! \int_{-\ell/2}^{\ell/2} \! ds_2 
\nonumber\\
&&
\bigg\{  
\frac{z^2}{| \rv + \uv |^6} 
- \frac{6z(y_1 r + (s_2z-s_1))(y_2 r + (s_2-s_1z))}{| \rv + \uv |^8} 
\nonumber\\
&&\quad+
\frac{9 (y_1 r + (s_2z-s_1))^2 (y_2 r + (s_2-s_1z))^2}{| \rv + \uv |^{10}}
\bigg\}, 
\label{eq:fint1}
\ea
where $\rv$ is the separation between the centers of mass of the two polymers, and $\uv \equiv s_2 \tv^{(2)} - s_1 \tv^{(1)}$. The three parameters $y_1 \equiv \tv^{(1)}\!\cdot\!\rh$, $y_2 \equiv \tv^{(2)}\!\cdot\!\rh$, and $z \equiv \tv^{(1)}\!\cdot\!\tv^{(2)}$ characterize the orientational configuration of the pair of rod-like polymers, where $\rh \equiv \rv/r$, and $\tv^{(1)}$ and $\tv^{(2)}$ are unit vectors characterizing the polymers' orientations. The parameters $y_1$, $y_2$ and $z$ are not completely independent, which we see as follows. Take $\rh$ to be parallel to the reference axis in spherical coordinate geometry, and write $\theta$ for the polar angle and $\phi$ for the azimuthal angle (see Fig.~\ref{fig:orientations}). We then have $\tv^{(1)}=(\sin\theta_1 \cos\phi_1,\sin\theta_1 \sin\phi_1, \cos\theta_1)$, $\tv^{(2)}=(\sin\theta_2 \cos\phi_2,\sin\theta_2 \sin\phi_2, \cos\theta_2)$, and $\rh = (0,0,1)$. We obtain $y_1 = \tv^{(1)}\cdot\rh = \cos\theta_1$, $y_2 = \tv^{(2)}\cdot\rh= \cos\theta_2$, and $z = \tv^{(1)}\cdot\tv^{(2)} = \sqrt{1-y_1^2} \sqrt{1-y_2^2} \psi + y_1 y_2$, where $\psi \equiv \cos(\phi_2-\phi_1)$. The two rods are co-planar if $\phi_1=\phi_2 + q\pi$ (or $|\psi|=1$; $q = 0, \pm 1, \pm 2, \ldots$).

In what follows, we study the behavior of $F_0$ in two limiting regimes: (i)~the near-field regime, $r \ll \ell$, where we consider two {\em rods} that are so near each other that the length of each rod can practically be assumed to be infinite, and (ii)~the far-field regime, $r \gg \ell$. 

\subsubsection{Near-field regime}
\label{sec:polymers_random_near}

We first consider the case where the length $\ell$ of each rod is much greater than the separation between the rods,  $r\ll \ell$. The problem becomes analytically tractable if we consider the limiting case of infinitely long rods. For such a case, the shortest length separation vector $\Rv_{12}^\ast$ is perpendicular to both $\tv^{(1)}$ and $\tv^{(2)}$. As we show in App.~\ref{app:near-field-distinct}, $F_0$ can be in that case approximated by 
\be
F_0 =
-\frac{M^2 \alpha_0^2 \, k_{\mathrm{B}}T}{64 \pi (\epsilon\epsilon_0)^2 \ell^2} 
\frac{z^2}{\sqrt{1-z^2}}\frac{1}{|\Rv_{12}^\ast|^4},
\label{eq:F0_nearfield}
\ee
where 
$
\Rv_{12}^\ast = 
\rv + \frac{(y_1 z - y_2)r}{1 - z^2} \tv^{(2)} - \frac{(y_1 - y_2 z)r}{1 - z^2} \tv^{(1)}.
$
The interaction energy is {\em more attractive} if the pair of polymers are {\em more aligned} (i.e., larger $z$), and {\em vanishes} if the two polymers are {\em perpendicular} to each other (i.e., $z$ is zero). The result is invariant under $z \rightarrow -z$, which is to be expected as the polymers have inversion symmetry. 

The above form of the interaction energy is also fully consistent with the general form for the vdW interactions between two anisotropic rods in the non-retarded limit \cite{Rajter1} if one takes into account the form of the anisotropy considered in deriving Eq.~(\ref{eq:F0_nearfield}).

\subsubsection{Far-field regime}

Let us consider the far-field regime, i.e., the case of rods of length $\ell$ which are short compared to their distance of separation, $r\gg \ell$. As the polymer lengths are finite, we can rescale $s_1, s_2$ in units of $\ell$ so that $-1/2 < s_1,s_2 < 1/2$. From Eq.~(\ref{eq:fint1}), $F_0$ can be put in a form where the integrand is dimensionless:
\ba
&&F_0 = -\frac{M^2 \alpha_0^2 \, k_{\mathrm{B}}T}{32 \pi^2 (\epsilon\epsilon_0)^2 r^6} 
\int_{-1/2}^{1/2} \! ds_1 \! \int_{-1/2}^{1/2} \! ds_2 
\nonumber\\
&&
\bigg\{  
\frac{z^2}{|\rh+x \uv|^6} 
- \frac{6z(y_1+(s_2z-s_1)x)(y_2+(s_2-s_1z)x)}{|\rh+x \uv|^8} 
\nonumber\\
&&\quad+
\frac{9 (y_1+(s_2z-s_1)x)^2 (y_2+(s_2-s_1z)x)^2}{|\rh+x \uv|^{10}}
\bigg\}.
\label{eq:F_distinct_farfield}
\ea
Here $x \equiv \ell/r$, which is small in the far-field regime. Expanding in powers of $x$ and integrating over $s_1$ and $s_2$ (see App.~\ref{app:far-field-distinct}), we can put Eq.~(\ref{eq:F_distinct_farfield}) in the following form: 
  \ba
F_0 &=& -\frac{M^2 \alpha_0^2 \, k_{\mathrm{B}}T}{32 \pi^2(\epsilon\epsilon_0)^2 r^6} 
\Big\{ (z - 3 \, y_1y_2)^2 
\nonumber\\
&&
+
\Big[
\frac{3}{4} \big( y_1^2 + y_2^2 - 50 \, y_1^2y_2^2 + 60 \, y_1^2y_2^2(y_1^2+y_2^2) \big)
\nonumber\\
&&+y_1y_2 \big(18-35(y_1^2+y_2^2) \big) z 
\nonumber\\
&&
+ \frac{3}{4} \big( 9 \, (y_1^2+y_2^2)-2 \big) z^2
\Big] \frac{\ell^2}{r^2}
\Big\}.
\label{eq:F0_farfield}
\ea
The leading order term goes as $r^{-6}$, which is characteristic of interactions between pairs of isotropic particles. 

Note that $F_0$ is also invariant under the operations of $\tv^{(1)}\rightarrow -\tv^{(1)}$ (which changes the sign of both $y_1$ and $z$) and/or $\tv^{(2)}\rightarrow -\tv^{(2)}$ (which changes the sign of both $y_2$ and $z$). This is expected, as the free energy cannot depend on whether one defines one end or the other of a rod as its ``head" (or ``tail").   
 \begin{figure}
		\includegraphics[width=0.35\textwidth]{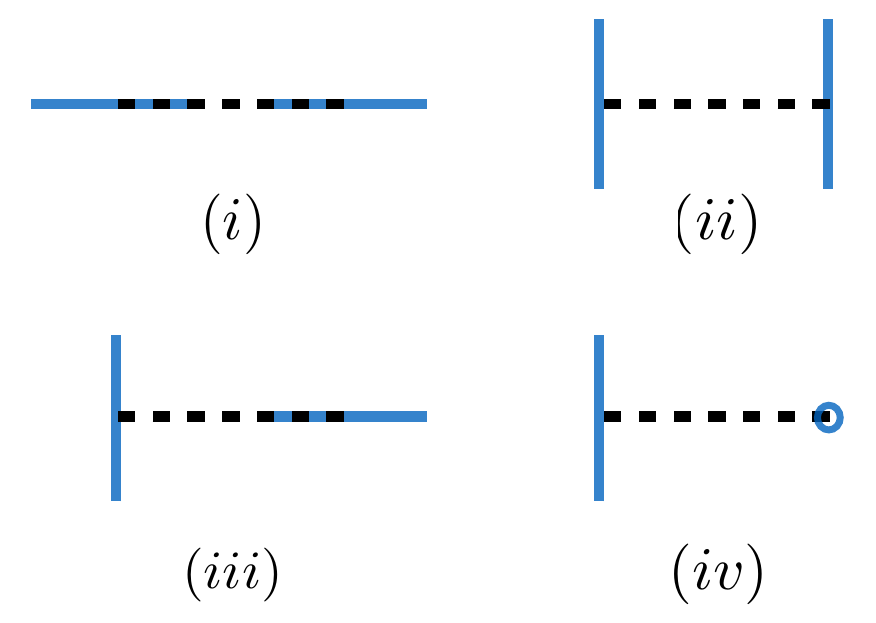}
	\caption{Four configurations of a pair of polarizable  rods: (i)~$|z|=|y_1|=|y_2|=1$, (ii)~$|z|=1, y_1=y_2=0$, (iii)~$z= y_1=0, |y_2|=1$, and (iv)~$z= y_1=y_2=0$. The interaction free energy is minimum for case~(i) (where rods are perfectly aligned with each other and with the vector joining their centers of mass), higher for cases~(ii) and (iii), and maximum for case~(iv) (where the rods are perpendicular to each other, and each rod is also perpendicular with the vector joining their centers of mass).} 
\label{fig:configs}
\end{figure}

We next consider $F_0$ for the following four illustrative cases (cf. Fig.~\ref{fig:configs}): ({\sl i})~$|z|=|y_1|=|y_2|=1$, ({\sl ii})~$|z|=1, y_1=y_2=0$, ({\sl iii})~$z= y_1=0, |y_2|=1$, and ({\sl iv})~$z= y_1=y_2=0$: 
\ba
F_0
&=& \left\{ \begin{array}{ll}
 -\frac{M^2 \alpha_0^2 \, k_{\mathrm{B}}T}{32 \pi^2(\epsilon\epsilon_0)^2 r^6}  \left( 4 + \frac{14\ell^2}{r^2} \right) &
   \mbox{($|z|=|y_1|=|y_2|=1$)},
   \vspace{3mm}\\
   -\frac{M^2 \alpha_0^2 \, k_{\mathrm{B}}T}{32 \pi^2(\epsilon\epsilon_0)^2 r^6}  \left( 1 - \frac{3\ell^2}{2r^2} \right) &
   \mbox{($|z|=1, y_1=y_2=0$)},
   \vspace{3mm}\\
-\frac{M^2 \alpha_0^2 \, k_{\mathrm{B}}T}{32 \pi^2(\epsilon\epsilon_0)^2 r^6} \left( \frac{3\ell^2}{4r^2} \right) &
   \mbox{($y_1 = z = 0, |y_2| = 1$)},
   \vspace{3mm}\\
0 &
   \quad \mbox{($z = y_1 = y_2 = 0$)},
   \end{array}  \right.
\ea
to the order of $O( r^{-10} )$.
$F_0$ is most negative for case~({\sl i}), i.e., attraction between the rods is strongest if they are aligned with each other, and each rod is also aligned with the vector joining their centers of mass. The interaction free energy is least negative for case~({\sl iv}), i.e., the rods are least attractive when the orientations of the rods and the vector joining their centers of mass are all perpendicular to one another. For a pair of rods that lie in the same plane as the vector joining their centers of mass, as both rods rotate about their own backbones perpendicular to the plane through their centers of mass, the interaction free energy interpolates between power law decays of the form $r^{-6}$ and $r^{-8}$. 

\subsection{Identical polymers }

Next, we consider the situation where the polymers are identical, viz., $\alpha^{(1)}(s) = \alpha^{(2)}(s) \equiv \alpha(s)$. Equation~(\ref{eq:Fint_simple}) then becomes 
\ba
&&F_{{\rm int}} = -\frac{k_{\mathrm{B}}T}{32\pi^2(\epsilon\epsilon_0)^2} \int  \! \frac{ds}{\mon}  \! \int  \! \frac{ds'}{\mon}  \,     
{\alpha}(s) \, {\alpha}(s') 
\nonumber\\
&&\quad\times
\left[
\frac{\tv^{(1)}\!\cdot\!\tv^{(2)}}{R_{12}^3}
-\frac{3(\tv^{(1)}\!\cdot\!\Rv_{12})(\tv^{(2)}\!\cdot\!\Rv_{12})}{R_{12}^5}
\right]^2,
\label{eq:Fint_identical}
\ea
where we recall that $\Rv_{12} \equiv \rv + s' \tv^{(2)} - s \, \tv^{(1)}$.
The sequence average is now given by 
\be
\llangle F_{{\rm int}} \rrangle = F_0 + \delta F,
\label{eq:F_sum}
\ee
where $\llangle \dots \rrangle$ again denotes the  average over the sequence probability distribution $P_{\alpha}$ and gives two contributions: the first, $F_0$, corresponds to the absence of correlations between the rods (which is the same quantity we calculated for a pair of distinct polymers)
\ba
&&F_0 = -\frac{k_{\mathrm{B}}T {\bar{\alpha}}^2}{32\pi^2(\epsilon\epsilon_0)^2} 
\int \! ds \! \int \! ds'     
\bigg\{
\frac{\tv^{(1)}\!\cdot\!\tv^{(2)}}{R_{12}^3}
\nonumber\\
&&\quad\quad\quad\quad-
\frac{3(\tv^{(1)}\!\cdot\!\Rv_{12})(\tv^{(2)}\!\cdot\!\Rv_{12})}{R_{12}^5}
\bigg\}^2,
\ea
while the second contribution $\delta F$ corresponds to correlation of two identical polarizability sequences, assumed to be short ranged, of the form of a delta function [cf. also Eq. (\ref{equ:aa_ii})]  
\ba
\label{eq:deltaF}
&&\delta F = -\frac{k_{\mathrm{B}}T {\bar{g}}g}{32\pi^2(\epsilon\epsilon_0)^2} 
\int \! ds  \,  
\bigg\{
\frac{\tv^{(1)}\!\cdot\!\tv^{(2)}}{R^3}
\nonumber\\
&&\quad\quad\quad\quad-
\frac{3(\tv^{(1)}\!\cdot\!\Rv)(\tv^{(2)}\!\cdot\!\Rv)}{R^5}
\bigg\}^2,   
\ea
where $\bar{g} \equiv g/\mon$ is the root mean square value of $\delta\alpha=\alpha(s)-\alpha_0$ per unit length of each monomer, and $\Rv \equiv \Rv^{(2)}(s) - \Rv^{(1)}(s) = \rv + s(\tv^{(2)} - \tv^{(1)})$. 
We have already calculated $F_0$ in the foregoing subsection.  We now calculate $\delta F$ in the same two limiting regimes of near field and far field. 

\subsubsection{Near-field regime}
\label{sec:identical_polymers_near_zero}
\begin{figure}
		\includegraphics[width=0.5\textwidth]{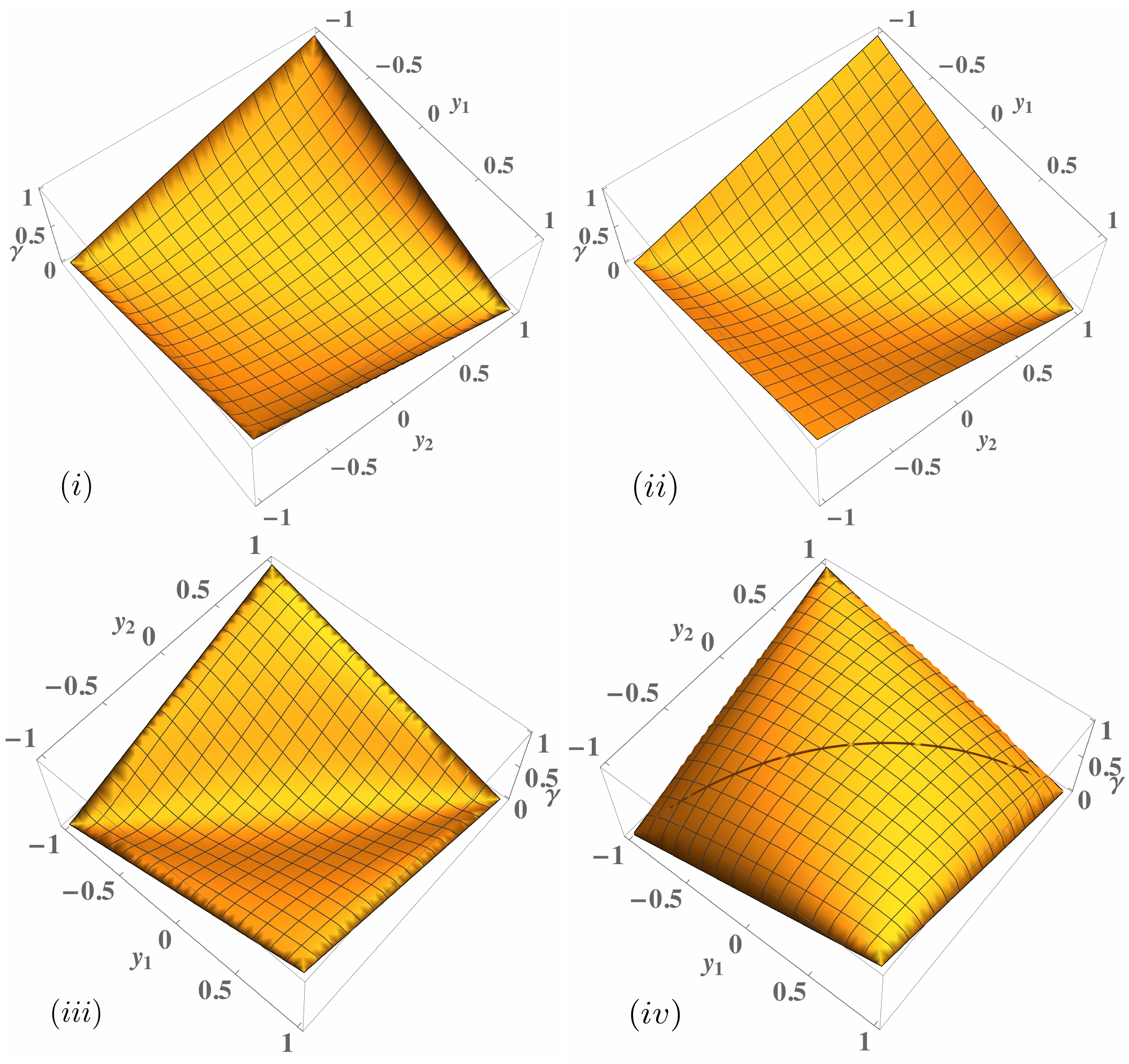}
	\caption{Behavior of $\gamma \equiv \frac{(y_1-y_2)^2}{2(1-z)}$ (vertical axis) for $-1 \leq y_1, y_2 \leq 1$ (the two horizontal axes) for the following four values of $\psi= \cos(\phi_2-\phi_1)$ (the cosine of the difference in azimuthal angles of the polymers defined with respect to $\rh$ as the reference axis): (i)~$\psi=-1$, (ii)~$\psi=0$, (iii)~$\psi=0.5$, and (iv)~$\psi=1$. We see that $0 \leq \gamma \leq 1$, and $\gamma$ has a maximum value of $1$ at $y_1=-y_2=\pm 1$, and a minimum value of $0$ at $y_1=y_2=\pm 1$.} 
\label{fig:gamma}
\end{figure}
Parallel to the near-field calculation for $F_0$ in Sec.~\ref{sec:polymers_random_near}, the evaluation of the integrals in $\delta F$ simplifies if we expand $\Rv$ about $\Rv^\ast$, the shortest length vector that connects the two polymers at equal arc-lengths ($s = s'$). Generally $\Rv^\ast \neq \Rv_{12}^\ast$; equality obtains only if $y_1=y_2=0$ (in which case $\Rv^\ast = \Rv_{12}^\ast = \rv$). 
In the near-field regime, we can make the approximation that the rods are infinitely long. Using Eq.~(\ref{eq:deltaF}), we find that $\delta F$ is given by (see App.~\ref{app:near-field-identical})
\be
\label{eq:deltaF_nearfield}
\delta F =
-\frac{3 k_{\mathrm{B}}T M g^2 \chi(y_1,y_2,z)}{16384 \sqrt{2} \pi (\epsilon\epsilon_0)^2 \ell r^5}, 
\ee
where the factor $\chi(y_1,y_2,z)$ encodes the orientation dependence and is defined by
\ba
&&\chi(y_1,y_2,z)=
\frac{1}{\sqrt{1-z} \left( 1 - \gamma \right)^{5/2}}
\bigg\{
9+14z+41z^2 
\nonumber\\
&&
- \frac{5(3+10z+3z^2)(y_1+y_2)^2}{1 - \gamma}
+\frac{105(y_1+y_2)^4}{4\left( 1 - \gamma \right)^2}
\bigg\}.\qquad
\label{eq:chi_orientation}
\ea
Here, $\gamma \equiv \frac{(y_1-y_2)^2}{2(1-z)}$ is a measure of the alignment or anti-alignment of a pair of identical rods. For $-1 \leq y_1, y_2, \psi \leq 1$ [where we recall that $\psi = \cos(\phi_2-\phi_1)$ is the cosine of the difference in azimuthal angles of the polymers defined with respect to $\rh$ as the reference axis], $\gamma$ varies between $0$ (for $y_1=y_2$) and $1$ (for $y_1=-y_2=\pm 1$ and $z=-1$). This behavior is plotted in Fig.~\ref{fig:gamma}.

We make the following observations. First, we note that $\delta F$ decays as $r^{-5}$, which is distinct from the $r^{-4}$ scaling of $F_0$ in the near-field regime. The $r^{-5}$ scaling is akin to a van der Waals interaction between an atom and a long thin rod. This is so because integrating along the lengths of two polymers subject to the constraint that $s=s'$ is equivalent to an integration along the length of one polymer. 

Next, $\delta F$ in Eq.~(\ref{eq:deltaF_nearfield}) depends on $z$, in contrast to the $z^2$-dependence of $F_0$. For polymers with uniform polarizabilities (which is what $F_0$ effectively describes), the interaction energy does not differentiate between the ``head" or ``tail" of a polymer, i.e., there is inversion symmetry (i.e., symmetry under the operation $\tv^{(1)} \rightarrow -\tv^{(1)}$, or $\tv^{(2)} \rightarrow -\tv^{(2)}$, or $z \rightarrow -z$). On the other hand, the breaking of inversion symmetry by $\delta F$ reflects the segment specificity of the polymer polarizability.

The behavior of $\delta F$ is correspondingly enriched by the ``head-tail" distinction of identical polymers. Compared to $F_0$ for which a pair of parallel rods and a pair of anti-parallel rods are degenerate, the value of $\delta F$ depends  on $\gamma$ whose value changes depending on whether rods are aligned or anti-aligned. Let us consider the behavior of $\delta F$ for six orientation configurations. The first four are shown in Fig.~\ref{fig:configs}, whilst the last two are $z = -1, y_1 = 1, y_2 \rightarrow -1$ and $z=-1, y_1 = y_2 = 0$ (which are the anti-parallel analogues of $y_1=y_2=1, z \rightarrow 1$ and $y_1=y_2=0, z \rightarrow 1$, respectively). To study the behavior of $\delta F$ for these configurations, we have to first determine the corresponding behavior of $\chi(y_1,y_2,z)$~\cite{footnote_configs}:
\ba
\chi(y_1,y_2,z)
&& \left\{ \begin{array}{ll}
 \rightarrow \frac{164}{\sqrt{1-z}} &
   \mbox{($y_1=y_2=1, z \rightarrow 1$)}
   \vspace{3mm}\\
   \rightarrow \frac{64}{\sqrt{1-z}} &
   \mbox{($y_1=y_2=0, z \rightarrow 1$)}
   \vspace{3mm}\\
= 336 \sqrt{2} &
   \mbox{($y_1=z=0, |y_2|=1$)}
   \vspace{3mm}\\
= 9 &
   \quad \mbox{($z = y_1 = y_2 = 0$)}
   \vspace{3mm}\\
\rightarrow  \frac{36}{\sqrt{2}(1-\gamma)^{5/2}}&
   \quad \mbox{($z = -1, y_1 = 1, y_2 \rightarrow -1$)}   
      \vspace{3mm}\\
= \frac{36}{\sqrt{2}} &
   \quad \mbox{($z=-1, y_1 = y_2 = 0$)}
   \end{array}  \right.
   \label{eq:compare}
\ea 
Applying Eq.~(\ref{eq:compare}) to Eq.~(\ref{eq:deltaF_nearfield}), we see that whereas $F_0$ was zero for perpendicular rods ($z=0$), $\delta F$ is negative and thus \emph{attractive} for the same configuration. 
We also see that out of the first four configurations, $\delta F$ is most attractive for perfect alignment (i.e., $y_1=y_2=z=1$), and least attractive when the rods are completely orthogonal (i.e., $y_1=y_2=z=0$). For sufficiently \emph{parallel} polymers ($z \rightarrow 1$), $\chi$ diverges as $(1-z)^{-1/2}$. 
Finally, each of the two anti-parallel configurations ($z = -1, y_1 = 1, y_2 \rightarrow -1$ and $z=-1, y_1 = y_2 = 0$) is not degenerate with its corresponding parallel counterpart (i.e., $y_1=y_2=1, z \rightarrow 1$ or $y_1=y_2=0, z \rightarrow 1$). Where $\chi$ for the parallel-rod configurations diverges as $(1-z)^{-1/2}$ as $z \rightarrow 1$, the value of $\chi$ for the analogous anti-parallel (i.e., $z=-1$) configurations can either diverge or not at all: in the case of $z = -1, y_1 = 1, y_2 \rightarrow -1$ (polymers sufficiently anti-parallel and the separation between their centers of mass is sufficiently parallel with the tangent vector of each polymer), $\chi$ diverges, but with an intriguing $(1-\gamma)^{-5/2}$ form as $\gamma \rightarrow 1$; on the other hand, the sixth configuration ($z=-1, y_1 = y_2 = 0$) has a finite and much lower value of $\delta F$ compared to the second configuration.

Thus at sufficiently short separations, ``molecular recognition" of polymers will act in the following \emph{two} senses: \emph{first}, the $r^{-5}$ scaling behavior of $\delta F$ becomes sufficiently strong and dominates over the $r^{-4}$ scaling of $F_0$, leading to pairs of identical polymers being selected over distinct ones; \emph{second}, amongst the pairs of identical polymers that are selected, the ones that are sufficiently parallel ($z \approx 1$) or anti-parallel ($z\approx -1$) with $\gamma \approx 1$ have the highest Boltzmann weight. 

\begin{figure}
		\includegraphics[width=0.45\textwidth]{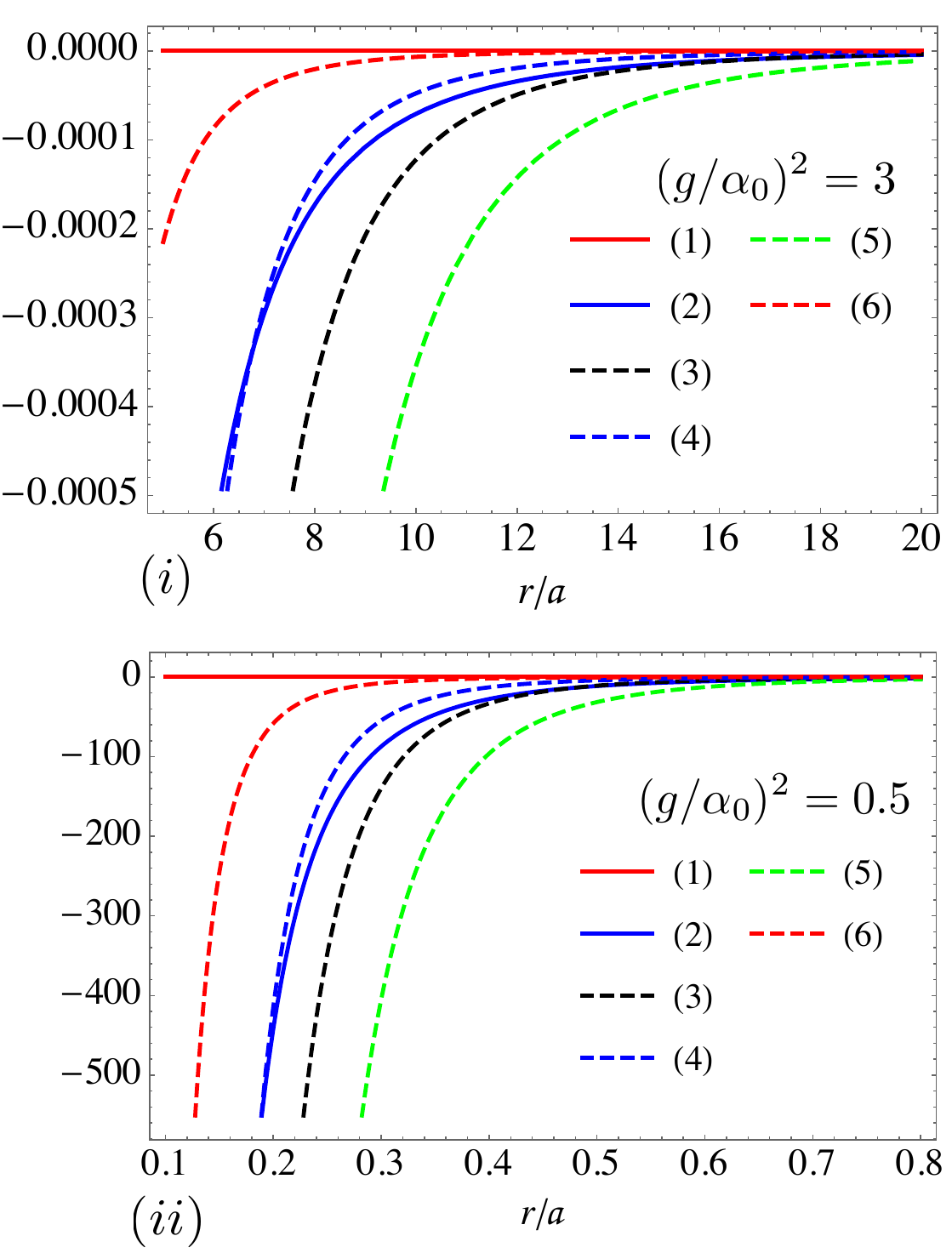}
	\caption{Comparison of the relative strengths of rescaled interaction free energies $f_0$ (solid line) and $\delta f$ (broken line) [defined in Eqs.~(\ref{eq:f0_deltaf})] for (i)~$g/\alpha_0 = 3$ and (ii)~$g/\alpha_0 = 0.5$, for each of the following four orientation configurations: (1) $y_1=z=0, y_2=1$ and $y_1=y_2=z=0$, (2) $y_1=y_2=z=1$ and $y_1=y_2=0, z=1$, (3) $y_1=y_2= z=1$, (4) $y_1=y_2=0$ and $z=1$, (5) $y_1=z=0$ and $y_2=1$ and (6) $y_1=y_2=z=0$. Here $\mon$ is the monomer size.} 
\label{fig:compare}
\end{figure}

What is the crossover length scale at which $\delta F$ begins to dominate over $F_0$? Such a length scale can be found by comparing the relative strengths of $\delta F$ and $F_0$. Using Eqs.~(\ref{eq:F0_nearfield}) and (\ref{eq:deltaF_nearfield}), we have
\be
\frac{\delta F}{F_0} = \frac{3 g^2 \tau(y_1,y_2,z) \ell}{256 \sqrt{2} M \alpha_0^2 r} ,
\label{eq:comparet}
\ee
where the orientation-dependent factor $\tau(y_1,y_2,z)$ is given by 
\ba
&&\tau(y_1,y_2,z) \equiv \frac{\sqrt{1+z}}{z^2(1-\gamma)^{5/2}}
\bigg\{
9+14z+41z^2 
\nonumber\\
&&
- \frac{5(3+10z+3z^2)(y_1+y_2)^2}{1 - \gamma}
+\frac{105(y_1+y_2)^4}{4\left( 1 - \gamma \right)^2}
\bigg\}.\qquad
\ea
Denoting by $R_c$ the length scale at which $\delta F/F_0 \sim 1$, we see from Eq.~(\ref{eq:comparet}) that $R_c$ is given by 
\be
R_c =  \frac{3 g^2 \tau(y_1,y_2,z)}{256 \sqrt{2} M \alpha_0^2} \ell. 
\ee
The length scale $R_c$ thus scales quadratically with $g/\alpha_0$. For rods that are perpendicular ($z = 0$), $R_c$ diverges, and thus $\delta F$ always dominates $\llangle F_{{\rm int}} \rrangle$. The behavior is shown in Fig.~\ref{fig:compare}, where we define 
\begin{subequations}
\ba
f_0 &\equiv& - \frac{z^2 a^4}{\sqrt{1-z^2}|\Rv_{12}^\ast|^4},  
\\
\delta f &\equiv& -\frac{3g^2 \chi(y_1,y_2,z) a^5}{256\sqrt{2} \alpha_0^2 r^5},
\ea
\label{eq:f0_deltaf}
\end{subequations}
such that $F_0 = (\alpha_0^2/(64 \pi \beta (\epsilon\epsilon_0)^2 \sqrt{1-z} \, a^6)) f_0$ and $\delta F = (\alpha_0^2/(64 \pi \beta (\epsilon\epsilon_0)^2 \sqrt{1-z} \, a^6)) \delta f$. 

For orthogonal rods ($z=0$), $F_0=0$ and thus $\llangle F_{{\rm int}} \rrangle$ is always dominated by $\delta F$. On the other hand, for rods that are sufficiently aligned $\delta F$ dominates over $F_0$ at sufficiently short separations, with the crossover controlled by the ratio $(g/\alpha_0)^2$. To compare $\delta F$ for different orientation configurations, we should multiply the curves by a factor $(1-z)^{-1/2}$, from which we see that aligned rods are more attracted to one another than unaligned ones. 

To see what the crossover means in practice, let us consider a dsDNA rod that is ten base-pairs long ($\ell = 10\mon$).  For the $z \rightarrow 1$ configurations, Fig.~\ref{fig:compare} then indicates that $\delta F$ already starts to dominate over $F_0$  at separations of order $\ell$ if the polarizability variance is three times the squared mean polarizability, but the crossover only occurs at separations of order $0.1\ell$ (i.e., the monomer size) if the polarizability variance is half the squared mean polarizability. Pairs of identical rods are thus more easily ``selected" over distinct rods if the identical rods each have a more heterogeneous polarizability sequence. 

\subsubsection{Far-field regime}
Next, we consider the case where the separation distance between the polymers is much larger than the length of each polymer, i.e., $r \gg \ell$. In this case, $\delta F$ can be expanded in powers of $\ell/r$. Correct to quadratic order in $\ell/r$, we obtain (see App.~\ref{app:far-field-identical})
\be
\label{eq:deltaF_farfield}
\delta F = -\frac{M g^2 k_{\mathrm{B}}T}{32\pi^2(\epsilon\epsilon_0)^2 r^6}      
\bigg\{
h_0 + \frac{h_2}{12} \left( \frac{\ell}{r} \right)^2 + O((\ell/r))^3 
\bigg\},
\ee
where the coefficients $h_0$ and $h_2$ are defined in Eqs.~(\ref{eq:h_coeffs}). 
From the leading order term, we see that $\delta F$ is most attractive when the pair of identical polymers are aligned ($z=-y_1=-y_2=1$) or anti-aligned ($z=-1, y_1=-y_2=1$ or $z=-1, y_1=-y_2=-1$). The leading order term is invariant under the inversion $\tv^{(1)} \rightarrow -\tv^{(1)}$, or $\tv^{(2)} \rightarrow -\tv^{(2)}$, or $z \rightarrow -z$, and the inversion symmetry is broken only at sub-leading order $O((\ell/r)^2)$. 

To leading order, we then remain with 
\be
\llangle F_{{\rm int}} \rrangle = -\frac{k_{\mathrm{B}}T(M^2 \alpha_0^2+ M g^2)(z - 3 \, y_1y_2)^2}{32 \pi^2(\epsilon\epsilon_0)^2 r^6} + O((\ell/r)^2),   
\label{eq:deltaF_far}
\ee
i.e., the only effect of the identity of polymers is to renormalize the overall sequence-averaged free energy. This seems again reasonable for the case considered, i.e., the far field regime.

\subsection{Identical polymers with non-zero sequence polarizability correlation length}

As in the previous case, we have $\alpha^{(1)}(s) = \alpha^{(2)}(s) \equiv \alpha(s)$, while in addition the correlator decays over a nonzero length scale, Eq.~(\ref{equ:beruwy}). The interaction free energy before sequence averaging is given by Eq.~(\ref{eq:Fint_identical}) and after sequence averaging we can express the free energy as the sum
\be
\llangle F_{{\rm int}} \rrangle = F_0 + \delta F, 
\ee
where the average $\llangle \dots \rrangle$ again denotes the sequence average, i.e., an average over the sequence probability distribution $P_{\alpha}$. Above,  $F_0$ is the part that does not involve sequence correlations, and $\delta F$ is the part that does involve such correlations. 

As before $F_0$ has already been calculated [see case~(i)]. The $\delta F$ contribution is given in complete analogy with the case of short range correlations, except that $\delta(s - s')$ is substituted by a Gaussian $f(s - s') \equiv ({\sqrt{2\pi}\sigma})^{-1} e^{-\frac{(s-s')^2}{2\sigma^2}}$. This leads to
\ba
\label{eq:deltaF_finite}
&&\delta F 
= -\frac{k_{\mathrm{B}}T g^2 }{32\pi^2(\epsilon\epsilon_0)^2 } 
\int_{-\ell/2}^{\ell/2}  \! \frac{ds_1}{\mon}  \! 
\int_{-\ell/2}^{\ell/2}  \! \frac{ds_2}{\mon}  \, 
\frac{\mon}{\sqrt{2\pi}\sigma} 
\\
&&\qquad\quad\times
e^{-\frac{(s_1-s_2)^2}{2\sigma^2}}
\left\{
\frac{\tv^{(1)}\!\cdot\!\tv^{(2)}}{R_{12}^3}
-\frac{3(\tv^{(1)}\!\cdot\!\Rv_{12})(\tv^{(2)}\!\cdot\!\Rv_{12})}{R_{12}^5}
\right\}^2
\nonumber
\ea
In what follows, we explore the effects of having a nonzero sequence polarizability correlation length. We consider two situations: 1)~\emph{large} sequence correlation length ($\sigma \gg \ell$) and 2)~\emph{short} sequence correlation length ($\sigma \ll \ell$). In the former situation, we find that $F_0$ always gives the dominant contribution to $\llangle F_{{\rm int}} \rrangle$. In the latter situation, on the other hand, we will find the emergence of three qualitative scaling regimes for $\delta F$: a \emph{far-field} regime characterized by $r^{-6}$ scaling, an \emph{intermediate near-field} regime characterized by $r^{-5}$ scaling, and an \emph{extreme near-field} regime characterized by $r^{-4}$ scaling. 

\subsubsection{Large sequence correlation length, $\ell \ll \sigma$}

Let us first consider the case of large sequence correlation length. From Eq.~(\ref{eq:deltaF_finite}) we see that $\delta F$ becomes smaller as $\sigma$ becomes larger, vanishing as $\sigma \rightarrow \infty$. The sequence-averaged interaction free energy of the identical rods then becomes no different from the interaction free energy between two uniformly polarized rods [i.e., $F_0$ in Eq.~(\ref{eq:fint})]. This is so, because a larger sequence correlation length brings about a stronger preference for the polarizabilities on each rod to be alike, and the rod thus appears to be more uniformly polarized.

As before, we can consider the near-field ($r \ll \ell \ll \sigma$) and far-field ($\ell \ll r \ll \sigma$) behaviors. As $\ell \ll \sigma$, we can expand the Gaussian kernel $e^{-(s_1-s_2)^2/2\sigma^2} \simeq 1 - (s_1-s_2)^2/2\sigma^2$. In the near-field regime, we consider for simplicity the case where $y_1=y_2=0$, and as before we approximate the rods by infinitely long ones. We find to subleading order 
\ba
&&\delta F = -\frac{M g^2 k_{\mathrm{B}}T}{1024\sqrt{2\pi}\pi(\epsilon\epsilon_0)^2 \sqrt{1-z}(1+z)^{3/2} \ell \sigma r^4} 
\nonumber\\
&&\times 
\bigg\{ 3(1+z+2z^2+2z^3) 
- (\frac{9}{2}+z+z^2)\frac{r^2}{\sigma^2} \bigg\}.        
\ea
In the near-field regime, $\delta F$ has the same leading order $r^{-4}$ scaling behavior as $F_0$, and the subleading contribution is of order $r^2/\sigma^2$. 
A comparison with Eq.~(\ref{eq:F0_nearfield}) shows that the leading order contribution of $\delta F$ differs from $F_0$ by a factor 
\ba
&&\frac{\delta F}{F_0} = \frac{3(1+z+2z^2+2z^3)\left\{ 1 - \frac{(y_1-y_2)^2}{2(1-z)} \right\}^2}{16 \sqrt{2\pi} z^2} \frac{g^2 \ell}{M \alpha_0^2 \sigma} 
\nonumber\\
&&\quad\quad+ O\left(\frac{r^2}{\sigma^2}\right).
\ea
Thus, in the near-field regime, the effect of $\delta F$ is to renormalize the prefactor of the vdW interaction energy, whilst the $r^{-4}$ scaling behavior remains essentially unchanged. As $M \sigma/\ell \rightarrow \infty$, $\llangle F_{{\rm int}} \rrangle$ becomes dominated by $F_0$. 

In the far-field regime ($\ell \ll r$), the ratio $x = \ell/r$ is small. Recalling Eq.~(\ref{eq:R12}), $\Rv_{12} = \rv + \uv$ where $\uv = s_2\tv^{(2)} - s_1\tv^{(1)}$, and rescaling $s_1, s_2, \sigma \rightarrow s_1 \ell, s_2\ell, \sigma\ell$, we have [analogous to Eq.~(\ref{eq:F_distinct_farfield})]
\ba
&&\delta F = -\frac{M^2 g^2 k_{{\mathrm B}} T}{32\pi^2(\epsilon\epsilon_0)^2 r^6}
\int_{-1/2}^{1/2} \!\!\!\!\!\! ds_1 \! \int_{-1/2}^{1/2} \!\!\!\!\!\! ds_2 \frac{a}{\sqrt{2\pi}\sigma\ell}e^{-\frac{(s_1-s_2)^2}{2\sigma^2}}
\nonumber\\
&&
\quad\times
\bigg\{  
\frac{z^2}{|\rh+x \uv|^6} 
- \frac{6z(y_1+(s_2z-s_1)x)(y_2+(s_2-s_1z)x)}{|\rh+x \uv|^8} 
\nonumber\\
&&\quad+
\frac{9 (y_1+(s_2z-s_1)x)^2 (y_2+(s_2-s_1z)x)^2}{|\rh+x \uv|^{10}}
\bigg\}.
\ea
For large correlation length, we can Taylor expand the Gaussian function to leading order. Then using Eq.~(\ref{eq:farfield}), we expand $\delta F$ to quadratic order in $x$: 
\ba
\delta F &\approx& -\frac{M g^2 k_{\mathrm{B}}T }{32\sqrt{2\pi}\pi^2(\epsilon\epsilon_0)^2 \sigma r^6} 
\int_{-1/2}^{1/2} \!\!\!\!\!\! ds_1 \! \int_{-1/2}^{1/2} \!\!\!\!\!\! ds_2  \,  
\nonumber\\
&&
\left(1-\frac{(s_1-s_2)^2}{2\sigma^2}\right)
(f_0 + f_1 x + f_2 x^2). 
\ea
The coefficients $f_i$ ($i=1,2,3$) are defined in Eqs.~(\ref{eq:f_coeffs}). 
Evaluating the integrals over $s_1$ and $s_2$ and restoring dimensions to $\sigma$ yields
\ba
\delta F &=& -\frac{M g^2 k_{\mathrm{B}}T \ell}{32\sqrt{2\pi}\pi^2(\epsilon\epsilon_0)^2 \sigma r^6} 
\\
&&\times\left\{ f_0 
+ \left( \frac{w_1+w_2}{12} \right) \frac{\ell^2}{r^2} - \frac{f_0}{12} \frac{\ell^2}{\sigma^{2}} \right\} 
\nonumber\\
&&+O(M\ell^5\sigma^{-3}r^{-8}).
\nonumber
\ea
The quantities $w_1$ and $w_2$ are defined in Eqs.~(\ref{eq:f_coeffs}). In the far-field regime, the contribution of $\delta F$ is always much smaller than that of $F_0$. The effect of the non-zero sequence correlation length is to make $\delta F$ less negative. 

\subsubsection{Short sequence correlation length, $\sigma \ll \ell$}
Next, we explore the consequences of having a short sequence correlation length. We can study three limiting cases: (i)~intermediate near-field ($\sigma \ll r \ll \ell$), (ii)~extreme near-field ($r \ll \sigma \ll \ell$), and (iii)~far-field ($\sigma \ll \ell \ll r$). 

We begin with case~(i), i.e.,  the \emph{intermediate near-field} behavior. Let us rescale $s_1$, $s_2$ and $\sigma$ in units of $r$; equation~(\ref{eq:deltaF_finite}) becomes
\ba
&&\delta F = -\frac{\bar{g} g \, k_{\mathrm{B}}T}{32\pi^2(\epsilon\epsilon_0)^2 r^5} 
\int_{-\ell/2r}^{\ell/2r} \!\!\!\!\!\! ds_1 \! 
\int_{-\ell/2r}^{\ell/2r} \!\!\!\!\!\! ds_2  
\frac{e^{-\frac{(s_1-s_2)^2}{2\sigma^2}}}{\sqrt{2\pi}\sigma} 
\nonumber\\
&&\quad\times
\bigg\{ 
\frac{z}{(1 + 2 (s_2 y_2 - s_1 y_1) + s_1^2 + s_2^2 - 2 z s_1 s_2)^{3/2}}
\nonumber\\
&&\quad\quad
- \frac{3(y_1 + s_2 z - s_1)(y_2 + s_2 - s_1 z)}{(1 + 2 (s_2 y_2 - s_1 y_1) + s_1^2 + s_2^2 - 2 z s_1 s_2)^{5/2}}
\bigg\}^2.
\nonumber\\
\ea
As the dimensionless quantities $\ell/r \gg 1$ and $\sigma \ll 1$, we can approximate the upper and lower limits of the integral over $u$ by $\infty$ and $-\infty$, respectively, and approximate the Gaussian kernel $(\sqrt{2\pi}\sigma)^{-1} \exp\big(-\frac{(s_1-s_2)^2}{2\sigma^2}\big)$ by the Dirac delta-function $\delta(s_1 - s_2)$. Physically this corresponds to the fact that at separations large compared to the sequence correlation length, the monomer polarizabilities on the same polymer appear to be uncorrelated. 
Performing the integration over $s_1$, we obtain the form of $\delta F$ in Eq.~(\ref{eq:deltaF}), whose near-field behavior has already been studied in Sec.~\ref{sec:identical_polymers_near_zero} and found to scale as $r^{-4}$.    

Next, let us turn to case~(ii), i.e., the \emph{extreme near-field} ($r \ll \sigma \ll \ell$) behavior. 
For simplicity, let us consider the case $y_1 = y_2 = 0$ whilst keeping $z$ arbitrary; this corresponds to a situation where the centers of the rods are directly opposite one another such that the vector joining these centers is perpendicular to the orientations of both rods; however, the rods are free to rotate in the same plane relative to each other. Defining $v \equiv s_2 - s_1$ and $u \equiv (s_1 + s_2)/2$, $\delta F$ becomes
 \ba
&&\delta F 
= 
-\frac{\bar{g} g \, k_{\mathrm{B}}T}{32\sqrt{2\pi}\pi^2(\epsilon\epsilon_0)^2 \sigma} 
\int_{-\ell/2}^{\ell/2} \!\!\!du 
\! \int_{-\ell}^{\ell} \!\!\! dv  \, 
 e^{-\frac{v^2}{2\sigma^2}}
 \nonumber\\
 &&\quad\quad\times
\bigg\{ 
\frac{z}{(r^2 + 2(1-z)u^2 + \frac{1}{2}(1+z)v^2)^{3/2}}
\nonumber\\
&&\quad\quad
- \frac{\frac{3}{4} (1+z)^2 v^2 - 3(1-z)^2 u^2}{(r^2 + 2(1-z)u^2 + \frac{1}{2}(1+z)v^2)^{5/2}}
\bigg\}^2.
\label{eq:deltaF_finite0}
\ea
We can scale out the dependence on $1-z$ by changing $u \rightarrow u/\sqrt{2(1-z)}$ and $v \rightarrow \sqrt{2}v/\sqrt{1+z}$:
\ba
&&\delta F 
= 
-\frac{\bar{g} g \, k_{\mathrm{B}}T}{32\sqrt{2\pi}\pi^2(\epsilon\epsilon_0)^2 \sqrt{1-z^2} \sigma} 
\\
 &&\quad\times
\int_{-\sqrt{\frac{1-z}{2}}\ell}^{\sqrt{\frac{1-z}{2}}\ell} \!\! du 
\! \int_{-\sqrt{\frac{1+z}{2}}\ell}^{\sqrt{\frac{1+z}{2}}\ell} \!\!  dv  \, 
 e^{-\frac{v^2}{(1+z)\sigma^2}}
 \nonumber\\
 &&\quad\times
\bigg\{ 
\frac{z}{(r^2 + u^2 +v^2)^{3/2}}
- \frac{\frac{3}{2} ((1+z)v^2 - (1-z)u^2)}{(r^2 + u^2 + v^2)^{5/2}}
\bigg\}^2.
\nonumber
\ea
After scaling $u$ and $v$ in units of $\sigma$ and defining $\lambda \equiv r/\sigma$, the equation above becomes 
 \ba
&&\delta F 
= 
-\frac{\bar{g} g \, k_{\mathrm{B}}T}{32\sqrt{2\pi}\pi^2(\epsilon\epsilon_0)^2 \sqrt{1-z^2} \sigma^5} 
 \nonumber\\
 &&\quad\quad\times
\int_{-\sqrt{\frac{1-z}{2}}\frac{\ell}{\sigma}}^{\sqrt{\frac{1-z}{2}}\frac{\ell}{\sigma}} \!\! du 
\! \int_{-\sqrt{\frac{1+z}{2}}\frac{\ell}{\sigma}}^{\sqrt{\frac{1+z}{2}}\frac{\ell}{\sigma}} \!\!  dv  \, 
 e^{-\frac{v^2}{1+z}}
\bigg\{ 
\frac{z}{(\lambda^2 + u^2 + v^2)^{3/2}}
\nonumber\\
&&\quad\quad\quad\quad
- \frac{\frac{3}{2} ((1+z)v^2 - (1-z)u^2)}{(\lambda^2 + u^2 + v^2)^{5/2}}
\bigg\}^2.
\label{eq:deltaF_extreme_nearfield}
\ea
As $\sigma / \ell \ll 1$, we can replace the upper (lower) limit of each integral by $\infty$ ($-\infty$). 
For simplicity, we focus on the behavior for $z \rightarrow 1$. 
For small $r/\sigma$, we find (see App.~\ref{app:extreme_near_field}) 
\be
\delta F \rightarrow -\frac{249 M g^2 k_{\mathrm{B}}T}{2048 \pi^{3/2} (\epsilon\epsilon_0)^2 \sqrt{1-z} \ell \sigma r^4}
\qquad (z \rightarrow 1).
\label{eq:extreme_near_field}
\ee
Used in conjunction with Eq.~(\ref{eq:F0_nearfield}), we find 
\be
\frac{\delta F}{F_0} \rightarrow \frac{249 \sqrt{2} g^2 \ell}{32 \sqrt{\pi} M \alpha_0^2\sigma} \qquad (z \rightarrow 1).
\label{eq:extreme_near_field_comparison}
\ee
Thus, $\delta F$ has the same scaling behavior as $F_0$ for $r \ll \sigma \ll \ell$. In contradistinction to $\delta F$ as obtained for rods of zero sequence correlation length (see Sec.~\ref{sec:identical_polymers_near_zero}), $\delta F$ of rods of nonzero sequence correlation length exhibits a crossover from $r^{-5}$ scaling to $r^{-4}$ scaling at $r \lesssim \sigma$, and $\delta F$ is larger than $F_0$ by a factor proportional to $(g^2/\alpha_0^2)(\ell / M \sigma)$.

Finally, we come to case~(iii), i.e., the \emph{far-field} ($\sigma \ll \ell \ll r$) behavior. To leading order in $r$, we can approximate Eq.~(\ref{eq:deltaF_finite}) by
\ba
&&\delta F = 
-\frac{M g^2 k_{\mathrm{B}}T \sigma (z-3y_1y_2)^2}{32\sqrt{2\pi}\pi^2(\epsilon\epsilon_0)^2 \ell r^6} 
\\
&&\quad\quad\quad\times
\int_{-\ell/2\sigma}^{\ell/2\sigma} \!\!\!\!\!\! ds_1 
\! \int_{-\ell/2\sigma}^{\ell/2\sigma} \!\!\!\!\!\!  ds_2  \, 
 e^{-\frac{(s_1-s_2)^2}{2}} + O(r^{-7})
 \nonumber\\
 &&=-\left(
 \frac{\sqrt{2 \pi} \ell
   }{\sigma} \text{erf}\left(\frac{\ell}{\sqrt{2}
   \sigma}\right)
    +2 e^{-\frac{\ell^2}{2
   \sigma^2}}-2
   \right)
   \nonumber\\
   &&\quad\quad\quad\times
   \frac{M g^2 k_{\mathrm{B}}T \sigma (z-3y_1y_2)^2}{32\sqrt{2\pi}\pi^2(\epsilon\epsilon_0)^2 \ell r^6}
   + O(r^{-7}).
 \nonumber
\ea
In the far-field regime, the only effect of having a nonzero sequence correlation length is a simple quantitative modification to the prefactor of $\delta F$, whose leading order scaling behavior and orientation dependence remain unchanged. At large separations, the sequence correlation length appears point-like if it is much shorter than the rod length. 
As expected, we recover the result Eq.~(\ref{eq:deltaF_far}) in the limit $\ell/\sigma \rightarrow \infty$.
 
\section{Summary and Discussion}

We have analyzed the van der Waals (vdW) interaction (free) energy between two rigid, rod-like heteropolymers with sequence specific polarizabilities assuming that each monomer in the sequence has a different, fixed (quenched) polarizability. Two fundamentally disjoint cases were treated. In the first case, we assumed that the polymers are distinct, each with a different polarizability sequence distribution.  In the second case, we assumed on the contrary that the interacting polymers are identical, with the same polarizability sequence distribution. Surprisingly, the results for the interaction energy in the two cases are fundamentally different. 

As a summary of our results, we find that in the near-field regime ($r \ll \ell$) for two distinct polymers, each having an average polarizability $\alpha_0$ and $M$ monomers, the interaction free energy has the form of Eq.~(\ref{eq:F0_nearfield}):
\be
F_0 =
-\frac{M^2 \alpha_0^2 \, k_{\mathrm{B}}T}{64 \pi (\epsilon\epsilon_0)^2 \ell^2} 
\frac{z^2}{\sqrt{1-z^2}}\frac{1}{|\Rv_{12}^\ast|^4}, 
\ee
where $\Rv^*_{12}$ is the shortest length separation vector between the rods, while in the far-field limit ($r \gg \ell$) the vdW interaction energy is in the form of Eq.~(\ref{eq:F0_farfield}):
\ba
F_0 &=& -\frac{M^2 \alpha_0^2 \, k_{\mathrm{B}}T (z - 3 \, y_1y_2)^2}{32 \pi^2(\epsilon\epsilon_0)^2 r^6}, 
\ea
where $r$ is the separation with $y_1 \equiv \tv^{(1)}\!\cdot\!\rv/r$, $y_2 \equiv \tv^{(2)}\!\cdot\!\rv/r$, and $z \equiv \tv^{(1)}\!\cdot\!\tv^{(2)}$. 

For identical polymers the interaction energy again decouples into the $F_0$ term, identical to above, and a correction $\delta F$ due to the polarization correlation between the two sequences. We studied two situations: (i)~the intra-chain correlation length $\sigma$ of the sequence distribution of polarizabilities is much larger than the polymer length $\ell$, and (ii)~$\sigma$ is much shorter than $\ell$. For the first situation, we found that the leading order contribution of $\delta F$ does not change the scaling behavior of the vdW interaction energy, and in the far-field regime the vdW interaction energy is always dominated by $F_0$. For the second situation, three qualitatively distinct scaling regimes emerge. In the \emph{far-field} regime ($\sigma \ll \ell \ll r$), we derive to the leading order
\be
\delta F  = -\frac{M g^2 k_{\mathrm{B}}T (z - 3 \, y_1y_2)^2}{32 \pi^2(\epsilon\epsilon_0)^2 r^6},   
\ee
which obviously just renormalizes the magnitude of the interaction energy in the case of two distinct polymers.
In the \emph{intermediate near-field} regime ($\sigma \ll r \ll \ell$), $\delta F$ exhibits $r^{-5}$ scaling, with the form of Eq.~(\ref{eq:deltaF_nearfield}): 
\be
\delta F =
-\frac{3 M g^2 k_{\mathrm{B}}T \chi(y_1,y_2,z)}{16384 \sqrt{2} \pi (\epsilon\epsilon_0)^2 \ell r^5}, 
\ee
where $\chi(y_1,y_2,z)$ is given by Eq.~(\ref{eq:chi_orientation}):
\ba
&&\chi(y_1,y_2,z)=
\frac{1}{\sqrt{1-z} \left( 1 - \gamma \right)^{5/2}}
\bigg\{
9+14z+41z^2 
\nonumber\\
&&\quad
- \frac{5(3+13z)(y_1+y_2)^2}{1 - \gamma}
+\frac{105(y_1+y_2)^4}{4\left( 1 - \gamma \right)^2}
\bigg\}
\ea
and $\gamma \equiv \frac{(y_1-y_2)^2}{2(1-z)}$. 
In the \emph{extreme near-field} regime ($r \ll \sigma \ll \ell$), $\delta F$ exhibits $r^{-4}$ scaling [see Eq.~(\ref{eq:extreme_near_field})], and is larger than $F_0$ by a factor proportional to $g^2 \ell / M \alpha_0^2 \sigma$ [see Eq.~(\ref{eq:extreme_near_field_comparison})]. 
Although the identity of sequences appears to simply renormalize the strength of the vdW interaction in the regime of large separations, in the regime of small separations such an identity generates a distinct angular dependence in the dominant order, and the interaction free energy even acquires a novel $r^{-5}$ scaling form in the interval $\sigma \ll r \ll \ell$. 

Without giving any calculational details, we also note that for a pair of Gaussian chains separated by a distance larger than their  radii of gyration $R_g^{(1)}$ and $R_g^{(2)}$, and which are characterizable by a Gaussian spatial distribution of monomers and an isotropic orientation distribution 
\ba
&&\overline{t_a^{(1)}(s) \, t_b^{(2)}(s')} = 0;
\\
&&\overline{t_a^{(1)}(s) \, t_b^{(1)}(s)} = \overline{t_a^{(2)}(s) \, t_b^{(2)}(s)} = \delta_{ab}
\ea
(where the overhead bars denote averaging over the monomer segment orientations in the high-temperature limit), the correlation of identical sequences only quantitatively renormalizes the prefactor of the vdW interaction energy and does not introduce any essential change to its scaling behavior. At large separations, $r \gg [(R_g^{(1)})^{2} + (R_g^{(2)})^{2}]^{1/2}$, the vdW interaction energy scales as $r^{-6}$, whilst at shorter separations (of the order of the radii of gyration) it scales as $r^{-4}$. 

We have demonstrated for a pair of rod-like polymers that the separation $R_c$ at which the correction due to polarization correlation, $\delta F$, becomes comparable to $F_0$ grows as $(g/\alpha_0)^2$. At separations smaller than $R_c$, $\delta F$ in fact dominates the vdW interactions between them. One can interpret this  as a {\em sequence recognition effect}, which would tend to separate interactions between pairs of identical heteropolymers from interactions between pairs of distinct heteropolymers. We intend to explore the various subtle consequences and generalizations of this effect in the near future. Compared to the case of near-field rod-like heteropolymers the sequence recognition effects are much smaller in all the other explored situations. 

\section{Acknowledgments}

BSL would like to thank G. Klimchitskaya and V. Mostepanenko for enlivening discussions.  BSL and RP would like to acknowledge the financial support of the Agency for research and development of Slovenia under grant  N1-0019. 

\appendix

\section{Derivation of $F_0$, Eq.~(\ref{eq:fint})}
\label{app1}
We denote the rod orientations by unit vectors $\tv^{(1)}$ and $\tv^{(2)}$, and the centers of masses of the rods by position vectors $\cv^{(1)}$ and $\cv^{(2)}$: 
\begin{subequations}
\ba
&&\Rv^{(1)}(s_1) = \cv^{(1)} + s_1 \tv^{(1)},
\\
&&\Rv^{(2)}(s_2) = \cv^{(2)} + s_2 \tv^{(2)}.
\ea
\end{subequations}
The vector linking one point $s_1$ on the first rod to a point $s_2$ on the second rod is given by 
\ba
\Rv_{12} &\equiv& \Rv^{(2)}(s_2) - \Rv^{(1)}(s_1)  
= \rv + \uv,  
\label{eq:R12}
\ea
where we have defined
\ba
\rv &\equiv& \cv^{(2)} - \cv^{(1)},
\\
\uv &\equiv& s_2 \tv^{(2)} - s_1 \tv^{(1)}.
\ea
The orientation of the vector joining the two centers of mass is specified by $\rh \equiv \rv/r$, and the orientational configuration of the pair of polymers can be specified by the three quantities: $y_1 \equiv \tv^{(1)}\!\cdot\!\rh$, $y_2 \equiv \tv^{(2)}\!\cdot\!\rh$, and $z \equiv \tv^{(1)}\!\cdot\!\tv^{(2)}$. From these quantities, we deduce the following relations: 
\ba
\tv^{(1)}\!\cdot\!\uv &=& s_2 z - s_1,
\\
\tv^{(2)}\!\cdot\!\uv &=& s_2 - s_1 z,
\\
\uv\!\cdot\!\rh &=& s_2 y_2 - s_1 y_1, 
\\
u^2 &=& s_1^2+s_2^2-2s_1s_2z.
\ea
Making use of Eq.~(\ref{eq:R12}), we obtain
\ba
&&
\frac{\tv^{(1)}(s_1)\!\cdot\!\tv^{(2)}(s_2)}{R_{12}^3}
-\frac{3(\tv^{(1)}(s_1)\!\cdot\!\Rv_{12})(\tv^{(2)}(s_2)\!\cdot\!\Rv_{12})}{R_{12}^5}
\nonumber\\
&&= 
\frac{z}{|\rv + \uv|^3} - \frac{3(y_1 r + (s_2z-s_1))(y_2 r +(s_2-s_1z))}{|\rv + \uv|^5}
\nonumber\\
\ea
Used in conjunction with Eq.~(\ref{eq:fint}), the above result leads to Eq.~(\ref{eq:fint1}).

\section{Distinct rods -- near-field calculation}
\label{app:near-field-distinct}

First, we show that the shortest length separation vector $\Rv_{12}^\ast$ is perpendicular to both $\tv^{(1)}$ and $\tv^{(2)}$. We determine the values $s_1^\ast$ and $s_2^\ast$ that specify the arc-length positions of the end-points of the vector $\Rv_{12}^\ast$, and calculate $F_0$ using $\Rv_{12}^\ast$ and shifted arc-length coordinates $\nu_1 \equiv s_1 - s_1^\ast$ and $\nu_2 \equiv s_2 - s_2^\ast$. 

First let us minimize $R_{12}^2$ with respect to $s_1$ and $s_2$. From Eq.~(\ref{eq:R12}), we have 
\be
R_{12}^2 = r^2 + 2(s_2y_2-s_1y_1) r + (s_1^2+s_2^2-2s_1s_2z).
\ee 
From the minimization condition we obtain the following solutions $s_1^\ast$ and $s_2^\ast$:
\ba
s_1^\ast &=& \frac{(y_1-y_2 z) r}{1-z^2}, 
\\
s_2^\ast &=& \frac{(y_1 z-y_2) r}{1-z^2}.
\ea
We then find that 
\ba
\Rv_{12}^\ast &=& \rv + s_2^\ast \tv^{(2)} - s_1^\ast \tv^{(1)} 
\\
&=& 
\rv + \frac{(y_1 z - y_2)r}{1 - z^2} \tv^{(2)} - \frac{(y_1 - y_2 z)r}{1 - z^2} \tv^{(1)}
\nonumber
\ea
and 
\be
(R_{12}^\ast)^2 = \left\{ 1 - \frac{y_1^2 + y_2^2 - 2 y_1 y_2 z}{1 - z^2} \right\} r^2.
\ee
It can be verified that $\Rv_{12}^\ast \cdot \tv^{(1)} = \Rv_{12}^\ast \cdot \tv^{(2)} = 0$. 
Next, let us define shifted coordinates $\nu_1$ and $\nu_2$: 
\begin{subequations}
\ba
&&\nu_1 \equiv s_1 - s_1^\ast = s_1 - \frac{(y_1 - y_2 z)r}{1 - z^2},
\\
&&\nu_2 \equiv s_2 - s_2^\ast = s_2 - \frac{(y_1 z - y_2)r}{1 - z^2}. 
\ea
\end{subequations}
We can express $\Rv_{12}$ in terms of the shortest length vector and shifted coordinates:
\be
\Rv_{12} = \Rv_{12}^\ast + \nu_2 \tv^{(2)} - \nu_1 \tv^{(1)}. 
\label{eq:shortexp}
\ee
Using Eqs.~(\ref{eq:green_function}), (\ref{eq:fint}) and (\ref{eq:shortexp}), we obtain the following expression for $F_0$:
\bw
\ba
F_0 &=& -\frac{k_{\mathrm{B}}T{\bar{\alpha}}^2}{32 \pi^2 (\epsilon\epsilon_0)^2} 
\int_{-\ell/2}^{\ell/2} \! ds_1 \! \int_{-\ell/2}^{\ell/2} \! ds_2 \,   
\bigg\{  \frac{\tv^{(1)}\!\cdot\!\tv^{(2)}}{|\Rv_{12}|^3} - \frac{3(\tv^{(1)}\!\cdot\!\Rv_{12})(\tv^{(2)}\!\cdot\!\Rv_{12})}{|\Rv_{12}|^5}
\bigg\}^2
\nonumber\\
&=& 
-\frac{k_{\mathrm{B}}T{\bar{\alpha}}^2}{32 \pi^2 (\epsilon\epsilon_0)^2} 
\int_{-\ell/2 - \frac{(y_1 - y_2 z)r}{1 - z^2}}^{\ell/2 - \frac{(y_1 - y_2 z)r}{1 - z^2}} \! d\nu_1 \! 
\int_{-\ell/2 - \frac{(y_1 z - y_2)r}{1 - z^2}}^{\ell/2 - \frac{(y_1 z - y_2)r}{1 - z^2}} \! d\nu_2 \,   
\bigg\{ 
\frac{z^2}{(|\Rv_{12}^\ast|^2 + \nu_1^2 + \nu_2^2 - 2\nu_1\nu_2 z)^3}
\nonumber\\
&&\quad\quad\quad\quad\quad\quad\quad\quad- \frac{ 6 z (z\nu_2-\nu_1) (\nu_2-z\nu_1) }{(|\Rv_{12}^\ast|^2 + \nu_1^2 + \nu_2^2 - 2\nu_1\nu_2 z)^4}
+ \frac{ 9 (z\nu_2-\nu_1)^2 (\nu_2-z\nu_1)^2 }{(|\Rv_{12}^\ast|^2 + \nu_1^2 + \nu_2^2 - 2\nu_1\nu_2 z)^5}
\bigg\}.
\label{eq:f_int_self_averaged}
\ea
\ew
We now take $\ell\rightarrow\infty$. The second and third terms vanish upon integration, and only the first term yields a non-zero contribution. We thus obtain Eq.~(\ref{eq:F0_nearfield}). 

\section{Distinct rods -- far-field calculation}
\label{app:far-field-distinct}

In the far-field regime, $x =\ell/r \ll 1$, and by rescaling $s \rightarrow s\ell$, we can make the following approximation,
\ba
|\rv + \uv|^{-2n} &\rightarrow& r^{-2n} |\rh + x \uv|^{-2n}
\\
&\simeq& r^{-2n} \big( 1 - 2n \rh\!\cdot\!\uv \, x 
\nonumber\\
&&
+ (2n(n+1)(\rh\!\cdot\!\uv)^2 - nu^2) x^2 \big).
\nonumber
\ea
For small $x$, the integrand $\{\ldots\}$ in Eq.~(\ref{eq:F_distinct_farfield}) can be expanded: 
\be
\{ \ldots \} \simeq f_0 + f_1 \, x + f_2 \, x^2.
\label{eq:farfield}
\ee
Here, the coefficients of $x^n$ are given by $f_n$ ($n=1,2,3$):
\begin{subequations}
\ba
f_0 &\equiv& z^2 - 6y_1 y_2 z + 9y_1^2 y_2^2,
\\
f_1 &\equiv& v_1 s_1 + v_2 s_2,
\\
f_2 &\equiv& w_1 s_1^2 + w_2 s_2^2 + w_3 s_1 s_2
\ea
where
\ba
v_1 &\equiv&  6(3y_1y_2-z) ((5y_1^2-1)y_2 - 2y_1 z),
\\
v_2 &\equiv& 6(3y_1y_2-z) (2y_2 z - (5y_2^2 - 1)y_1)
\ea
and
\ba
w_1 &\equiv& 3[3(1-25y_1^2+60y_1^4)y_2^2 
\nonumber\\
&&-4y_1y_2(35y_1^2-9)z+3(9y_1^2-1)z^2],
\\
w_2 &\equiv& 3[3(1-25y_2^2+60y_2^4)y_1^2
\nonumber\\
&&-4y_1y_2(35y_2^2-9)z+3(9y_2^2-1)z^2],
\\
w_3 &\equiv& 3[4z^3 -76 y_1 y_2 z^2 
\nonumber\\
&&+ (2-22(y_1^2+y_2^2)+310y_1^2y_2^2)z 
\nonumber\\
&&
-360 y_1^3 y_2^3 + 12y_1y_2(5(y_1^2+y_2^2)-1)].
\ea
\label{eq:f_coeffs}
\end{subequations}
Integration over $s_1$ and $s_2$ yields
\ba
&&\int_{-1/2}^{1/2}ds_1\int_{-1/2}^{1/2}ds_2 \, f_0 = (z - 3 \, y_1y_2)^2,  
\\
&&\int_{-1/2}^{1/2}ds_1\int_{-1/2}^{1/2}ds_2 \, f_1 = 0, 
\\
&&\int_{-1/2}^{1/2}ds_1\int_{-1/2}^{1/2}ds_2 \, f_2 
\\
&&\quad= 
\frac{3}{4} [y_1^2 + y_2^2 - 50 \, y_1^2y_2^2 + 60 \, y_1^2y_2^2(y_1^2+y_2^2)]
\nonumber\\
&&\quad\quad+y_1y_2[18-35(y_1^2+y_2^2)]z 
\nonumber\\
&&\quad\quad+ \frac{3}{4}[9 \, (y_1^2+y_2^2)-2]z^2.
\nonumber
\ea
Making use of the above results, Eq.~(\ref{eq:fint1}) can be put in the form Eq.~(\ref{eq:F0_farfield}).

\section{Identical rods -- near-field calculation}
\label{app:near-field-identical}

Let us first derive the shortest length vector $R^\ast$. The square separation $R^2$ is given by 
\be
R^2 = r^2 + 2(y_2-y_1) r s + 2(1-z)s^2. 
\label{eq:R2}
\ee
By minimizing $R^2$ with respect to $s$, we obtain 
\ba
s^\ast = \frac{y_1 - y_2}{2(1-z)}r.&&
\ea
This gives
\be
\Rv^\ast = \rv + \frac{(y_1 - y_2) r}{2(1-z)}(\tv^{(2)} - \tv^{(1)}),
\ee
and 
\be
(R^\ast)^2 = ( 1 - \gamma ) r^2,
\ee
where $\gamma \equiv \frac{(y_1-y_2)^2}{2(1-z)}$. 
$\Rv^\ast$ coincides with $\rv$ (the vector joining the centers of mass of the polymers) if $y_1=y_2=0$.  
$\Rv^\ast$ is \emph{not} perpendicular to $\tv^{(1)}$ or $\tv^{(1)}$, but it is perpendicular to $\tv^{(2)} - \tv^{(1)}$: 
\ba
\label{eq:y_1y_2r}
&&\Rv^\ast\!\cdot\! \tv^{(1)} = \Rv^\ast\!\cdot\! \tv^{(1)} = \frac{1}{2}(y_1 + y_2)r, 
\\
&&\Rv^\ast\!\cdot\! (\tv^{(2)} - \tv^{(1)}) = 0.
\ea
Let us express
\be
\Rv = \Rv^\ast + \mu(\tv^{(2)} - \tv^{(1)}), 
\ee
where $\mu \equiv s - s^\ast$. Because of the orthogonality of $\Rv^\ast$ and $\tv^{(2)} - \tv^{(1)}$, we find 
\be
R^2 = |\Rv^\ast|^2 + 2(1-z)\mu^2 = (1-\gamma)r^2+2(1-z)\mu^2.  
\ee
Using the above results we have
\begin{subequations}
\ba
&&\tv^{(1)}\cdot\Rv = ry_1-(1-z)s
\nonumber\\
&&\quad
=\frac{1}{2}(y_1+y_2)r - (1-z)\mu,
\\
&&\tv^{(2)}\cdot\Rv = ry_2+(1-z)s
\nonumber\\
&&\quad
=\frac{1}{2}(y_1+y_2)r + (1-z)\mu,
\ea
\label{eq:t_project}
\end{subequations}
and 
\ba
&&\int_{-\ell/2}^{\ell/2} \!\!\!\!ds   
\left\{
\frac{\tv^{(1)}\!\cdot\!\tv^{(2)}}{R^3}
-\frac{3(\tv^{(1)}\!\cdot\!\Rv)(\tv^{(2)}\!\cdot\!\Rv)}{R^5}
\right\}^2
\nonumber\\
&&=
\int_{-\ell/2-\frac{(y_1-y_2)r}{2(1-z)}}^{\ell/2-\frac{(y_1-y_2)r}{2(1-z)}} \!\!d\mu   
\bigg\{
\frac{z}{((1-\gamma)r^2+2(1-z)\mu^2)^{3/2}}
\nonumber\\
&&\quad
-\frac{\frac{3}{4}(y_1+y_2)^2 r^2 - 3(1-z)^2 \mu^2}{((1-\gamma)r^2+2(1-z)\mu^2)^{5/2}}
\bigg\}^2.
\ea
Next we rescale $\mu \rightarrow \mu/\sqrt{2(1-z)}$ and let $\ell\rightarrow\infty$, obtaining
\ba
&&\int_{-\ell/2}^{\ell/2} \!\!\!\!ds   
\left\{
\frac{\tv^{(1)}\!\cdot\!\tv^{(2)}}{R^3}
-\frac{3(\tv^{(1)}\!\cdot\!\Rv)(\tv^{(2)}\!\cdot\!\Rv)}{R^5}
\right\}^2
\nonumber\\
&&=
\frac{1}{\sqrt{2(1-z)}}\int_{-\infty}^{\infty} \!\!\!\!\!\!d\mu   
\bigg\{
\frac{z}{((1-\gamma)r^2+\mu^2)^{3/2}}
\nonumber\\
&&\quad
-\frac{\frac{3}{4}(y_1+y_2)^2 r^2 - \frac{3}{2}(1-z) \mu^2}{((1-\gamma)r^2+\mu^2)^{5/2}}
\bigg\}^2
\nonumber\\
&&=
\frac{3\pi}{512\sqrt{2(1-z)}(1-\gamma)^{5/2}r^5} 
\nonumber\\
&&\quad \times
\bigg\{
9+14z+41z^2 
- \frac{5(3+10z+3z^2)(y_1+y_2)^2}{1-\gamma}
\nonumber\\
&&\quad\quad
+\frac{105(y_1+y_2)^4}{4(1-\gamma)^2}
\bigg\}
\ea
Used together with Eq.~(\ref{eq:deltaF}), we obtain 
\ba
&&\delta F =
-\frac{3 M g^2 k_{\mathrm{B}}T}{16384 \sqrt{2(1-z)} \pi (\epsilon\epsilon_0)^2 (1-\gamma)^{5/2} \ell r^5} 
\nonumber\\
&&\quad \times
\bigg\{
9+14z+41z^2 
- \frac{5(3+10z+3z^2)(y_1+y_2)^2}{1-\gamma}
\nonumber\\
&&\quad\quad
+\frac{105(y_1+y_2)^4}{4(1-\gamma)^2}
\bigg\},
\ea 
which is equivalent to Eq.~(\ref{eq:deltaF_nearfield}). 

\section{Identical rods -- far-field calculation}
\label{app:far-field-identical}

Using Eqs.~(\ref{eq:R2}) and (\ref{eq:t_project}), we can put Eq.~(\ref{eq:deltaF}) in the form
\ba
&&\delta F = -\frac{k_{\mathrm{B}}T {\bar{g}}g}{32\pi^2(\epsilon\epsilon_0)^2} 
\\
&&\quad \times \int_{-\ell/2}^{\ell/2}  \!\!\!\! \! ds 
\bigg\{
\frac{z^2}{r^6(1+2(y_2-y_1)(s/r)+2(1-z)(s/r)^2)^6}
\nonumber\\
&&\quad
-\frac{6z(y_1y_2r^2 + (1-z)(y_1-y_2)rs - (1-z)^2 s^2)}{r^8(1+2(y_2-y_1)(s/r)+2(1-z)(s/r)^2)^8}
\nonumber\\
&&\quad+ \frac{9(y_1y_2r^2 + (1-z)(y_1-y_2)rs - (1-z)^2 s^2)^2}{r^{10}(1+2(y_2-y_1)(s/r)+2(1-z)(s/r)^2)^{10}}
\bigg\}.
\nonumber
\ea
Let us expand the integrand in the above expression in powers of small $s/r$:
\ba
\delta F &=& -\frac{k_{\mathrm{B}}T {\bar{g}}g}{32\pi^2(\epsilon\epsilon_0)^2 r^6} \int_{-\ell/2}^{\ell/2}  \! ds \,     
\bigg\{
h_0 + h_1 \left( \frac{s}{r} \right) + h_2 \left( \frac{s}{r} \right)^2
\bigg\},
\nonumber
\ea
where 
\begin{subequations}
\ba
&&h_0 = (3 y_1 y_2 - z)^2,
\\
&&h_1 = 6 (y_1 - y_2) (1 + 10 y_1 y_2 - 3 z) (3 y_1 y_2 - z),
\\
&&h_2 = 3 (3 y_1^2 - 12 y_1 y_2 + 120 y_1^3 y_2 + 3 y_2^2 - 300 y_1^2 y_2^2 
   \nonumber\\
   &&\quad\quad
+ 
   660 y_1^4 y_2^2 + 120 y_1 y_2^3 
- 1320 y_1^3 y_2^3 + 660 y_1^2 y_2^4 
   \nonumber\\
   &&\quad\quad
+ 
   2 z - 38 y_1^2 z + 120 y_1 y_2 z - 408 y_1^3 y_2 z - 38 y_2^2 z 
   \nonumber\\
   &&\quad\quad+ 
   876 y_1^2 y_2^2 z - 408 y_1 y_2^3 z - 8 z^2 + 63 y_1^2 z^2 
      \nonumber\\
   &&\quad\quad
   - 
   164 y_1 y_2 z^2 + 63 y_2^2 z^2 + 6 z^3).
\ea
\label{eq:h_coeffs}
\end{subequations}
Now, performing the integration over $s$, the term linear in $s$ vanishes, and we obtain Eq.~(\ref{eq:deltaF_farfield}).

\section{Nonzero sequence correlation -- extreme near-field calculation}
\label{app:extreme_near_field}

The evaluation of $\delta F$ in Eq.~(\ref{eq:deltaF_extreme_nearfield}) for $z \rightarrow 1$ is aided by the use of the following integrals:
\ba
&&\int_{-\infty}^{\infty} \!\!\!\!\!du \! \int_{-\infty}^{\infty} \!\!\!\!\!dv
\frac{e^{-\frac{v^2}{1+z}}}{(\lambda^2+u^2+v^2)^3} 
= \frac{\pi e^{\frac{\lambda^2}{2(1+z)}}}{4(1+z)^2}
\\
&&\quad\quad\times
\left\{ K_0\!\left( \frac{\lambda^2}{2(1+z)} \right) 
+ \frac{1+z-\lambda^2}{\lambda^2} K_1\!\left( \frac{\lambda^2}{2(1+z)} \right) \right\},
\nonumber\\
&&\int_{-\infty}^{\infty} \!\!\!\!\!du \! \int_{-\infty}^{\infty} \!\!\!\!\!dv
\frac{v^2 e^{-\frac{v^2}{1+z}}}{(\lambda^2+u^2+v^2)^4} 
=\frac{5\pi^{3/2}}{32\lambda^4} 
U\!\left( \frac{3}{2}, -1, \frac{\lambda^2}{1+z} \right),
\nonumber\\
&&\int_{-\infty}^{\infty} \!\!\!\!\!du \! \int_{-\infty}^{\infty} \!\!\!\!\!dv
\frac{v^4 e^{-\frac{v^2}{1+z}}}{(\lambda^2+u^2+v^2)^5} 
=\frac{105\pi^{3/2}}{512\lambda^4}
U\!\left( \frac{5}{2}, -1, \frac{\lambda^2}{1+z} \right),
\nonumber
\ea
where $K_\nu(x)$ is a modified Bessel function of the second kind, and $U(a,b,x)$ is a confluent hypergeometric function. For small $\lambda$ we obtain the following leading order contributions: 
\ba
&&\int_{-\infty}^{\infty} \!\!\!du \! \int_{-\infty}^{\infty} \!\!\!dv
\frac{e^{-\frac{v^2}{1+z}}}{(\lambda^2+u^2+v^2)^3} 
= \frac{2\pi(1+z)^2}{\lambda^4},
\nonumber\\
&&\int_{-\infty}^{\infty} \!\!\!du \! \int_{-\infty}^{\infty} \!\!\!dv
\frac{v^2 e^{-\frac{v^2}{1+z}}}{(\lambda^2+u^2+v^2)^4} 
= \frac{\pi}{12\lambda^4},
\nonumber\\
&&\int_{-\infty}^{\infty} \!\!\!du \! \int_{-\infty}^{\infty} \!\!\!dv
\frac{v^4 e^{-\frac{v^2}{1+z}}}{(\lambda^2+u^2+v^2)^5} 
= \frac{\pi}{32\lambda^4}.
\ea


\begin{thebibliography}{99}
\bibitem{footnote_inversion_symmetry}
With a symmetric choice of arc-lengths range it is more straightforward to study the effects (or lack thereof) that the inversion of a polymer about its center of mass have on interactions between identical and distinct polarizability sequences.
\bibitem{footnote_configs}
For the first two configurations, we study the behavior for fixed values of $y_1=y_2$ and $z \neq 1$ as $z \rightarrow 1$, so $\gamma = 0$.
\end{thebibliography}
\end{document}